\documentclass[twocolumn]{aastex631}
\usepackage{amsmath}

\newcommand{\code}[1]{\texttt{#1}}

\graphicspath{{./}{Figures/}}

\begin{document}

\title{Evidence for a Sharp CO Snowline Transition in a Protoplanetary Disk \\
and Implications for Millimeter-wave Observations of CO Isotopologues}

\correspondingauthor{Chunhua Qi}
\email{cqi1@bu.edu}

\author[0000-0001-8642-1786]{Chunhua Qi}
\affiliation{Center for Astrophysics \textbar\, Harvard \& Smithsonian, 60 Garden St., Cambridge, MA 02138, USA}

\affiliation{Institute for Astrophysical Research, Boston University, 725 Commonwealth Avenue, Boston, MA 02215, USA}

\author[0000-0003-1526-7587]{David J. Wilner}
\affiliation{Center for Astrophysics \textbar\, Harvard \& Smithsonian, 60 Garden St., Cambridge, MA 02138, USA}

\begin{abstract}
Observations of CO isotopologue emission from protoplanetary disks at millimeter wavelengths are 
a powerful tool for probing the CO snowline, an important marker for disk chemistry, and also for
estimating total disk gas mass, a key quantity for planet formation. We use simple models to demonstrate
that the vertical thickness of an isothermal layer around the disk midplane has important effects on the CO column density radial profile, with a thick layer producing a sharp CO snowline transition. We simulate ngVLA and ALMA images 
to show that this sharp change in CO column density can be detected in the derivative of the radial profile 
of emission from optically thin CO isotopologue lines.
We apply this method to archival ALMA observations of the disk around the Herbig Ae star HD~163296 
in the \(\mathrm{C^{17}O}\) and \(\mathrm{C^{18}O}\) J=1-0 and J=2-1 lines 
to identify a sharp CO snowline transition near $\sim80$~au ($0\farcs8$ at 101 pc),
and show the CO column density decreases by more than a factor of 20.
This finding is consistent with previous inferences from the steep rise of N$_2$H$^+$ emission, 
which marks the location where CO depletes. 
We also demonstrate that the disk's thermal structure introduces significant systematic uncertainty to 
estimates of total disk gas mass derived from these lines. 
The substantial improvement in  sensitivity envisioned for the ngVLA over ALMA for observations of 
ground-state lines of CO isotopologues 
has the potential to extend this approach to a much larger population of disks. 

\end{abstract}


\section{Introduction} \label{sec:intro}

Planets form from dust and gas within protoplanetary disks surrounding young stars. 
In the cold disk midplane, temperatures are low enough that many molecules are frozen out 
onto grain mantles and depleted from the gas phase. 
The CO snowline, defined as the radius at which the midplane temperature is low enough for 
CO to freeze out onto dust grains, is an important feature of these disks. 
The CO snowline location is pivotal because of its effects on the chemical composition 
and physical properties of the disk. The locations of disk substructures relative to the snowlines 
(``condensation fronts'') of major volatile species, including H$_2$O, CO$_2$, CO, and N$_2$, 
are important for setting the chemistry in the disks \citep{Oberg+2023}.
Radial variations in the C/O ratio affect the composition of material accreted by forming planets, 
influencing the abundances in their atmospheres \citep[e.g.,][]{Oberg+2011}.
In addition, CO, and its less abundant isotopologues, are also commonly used as tracers 
of disk gas mass, essential for elucidating processes involved in planet building, as 
cold H$_2$ -- the primary mass constituent -- is impossible to observe directly. 
However, accurate conversion of CO emission to total gas mass requires accounting for 
CO that has frozen out of the gas phase, as well as photodissociation and potential 
chemical conversion processes \citep{Ansdell+2018,Bosman+2018,Krijt+2018}.

The cold midplane and its proximity to chemically active layers in disks can be probed directly
by observations of low excitation lines of trace species at millimeter wavelengths.  
The CO snowline is difficult to 
observe directly because of radial and vertical temperature gradients, leading to a warmer region
above the ``snow surface'' where CO remains in the gaseous state throughout \citep{Qi+2011}.
While emission from the most abundant main isotopologue $^{12}$CO in disks is bright, it is
also generally very optically thick and traces only an effective photospheric temperature 
high above the midplane. By contrast, emission from lower-abundance CO isotopologues 
($^{13}$CO, C$^{18}$O, C$^{17}$O, $^{13}$C$^{18}$O) have lower optical depths and emerge 
from deeper in the disk. The high sensitivity and resolution of ALMA for observing low-$J$ lines 
of these rarer CO isotopologues, and other trace species of cold gas in disks, has revolutionized 
studies of CO snowlines  \citep[e.g.,][]{Zhang+2017,Booth+2023}.
N$_2$H$^+$, a species whose abundance is greatly enhanced where freeze-out removes 
CO from the gas phase \citep{Bergin+2002} has shown promise as a chemical tracer of the CO snowline 
\citep{Qi+2013}, although the interpretation of N$_2$H$^+$is complicated by the 
three-dimensional structure 
of  the condensation front that arises from the details of the vertical temperature structure 
\citep{vantHoff+2017,Qi+2019}.
More directly, high-resolution observations of CO isotopologues, together with 
disk models, have been used to define the three-dimensional CO ``snow surface'' in disks, 
though this method is complicated by the fact that the most easily detected CO isotopologue lines 
can have substantial opacity \citep[e.g,][]{Pinte+2018}. 

The temperature structure of protoplanetary disks is set by a balance of heating and cooling that 
depends sensitively on the effects of grain growth, and size-dependent radial drift and vertical settling 
\citep[e.g.,][]{Dalessio+2006}. The interplay of these processes can have important effects on the 
location of the CO freeze-out region. For example, the models of \citet{Cleeves2016} predict that inward drift 
of millimeter-sized grains can result in a small radial temperature inversion that creates a second CO snowline. 
A more significant effect is predicted to stem from the relative scale heights of large and small grains that 
control the extent of a vertically isothermal region around the midplane (VIRaM) layer, where temperatures 
are nearly isothermal vertically at the CO freeze-out region. \citet{Qi+2019} showed that disks with a thick 
VIRaM layer exhibit bright, narrow N$_2$H$^+$ rings that peak in the slim radial zone between the CO and N$_2$ 
snowlines. While this chemical signature of a thick VIRaM layer is highly suggestive, direct evidence is still lacking.

In this paper, Section~\ref{sec:toy} presents a generic disk model to show that a thick VIRaM layer 
leads to a sharp drop in the CO column density across the CO snowline, by more than an order of magnitude, 
and that direct detection of such a ``sharp CO snowline transition'' can be made using observations of low
excitation rotational lines of rare CO isotopologues with high sensitivity, spatial resolution, and spectral resolution.
Section~\ref{sec:hd163296} applies the concepts from this model to archival ALMA observations of 
CO isotopologues from the large disk around the nearby Herbig Ae star HD~163296, 
hypothesized to possess a thick VIRaM layer from its N$_2$H$^+$ emission morphology, and provides 
strong evidence for the presence of a sharp CO snowline transition. 
Section~\ref{sec:discussion} discusses the implications of this thermal structure,
in particular for disk gas mass  estimates obtained from observations of CO isotopologues.  
In this context, the next generation Very Large Array \citep[ngVLA,][]{Murphy+2018} 
has the potential for substantially advancing this field as a result of its substantially 
higher sensitivity than ALMA for observations of rare CO isotopologue ground-state lines.

\section{Disk Vertical Thermal Structure Effects}
\label{sec:toy}
To investigate the impact of the disk's vertical thermal structure on the CO radial column density profile,
we employ a simple parametric disk model similar to that used by \citet{Andrews+2009}.
We then  simulate observations of the J=1--0 and J=2--1 line emission of rare CO isotopologues
to demonstrate the observable consequences of the VIRaM layer thickness.

\subsection{Disk Model Prescriptions}

We adopt a disk surface density ($\Sigma(r)$) that follows a power-law profile with an exponential taper
and a characteristic radius ($r_c$):
\begin{equation}
\Sigma(r) = \Sigma_0 \left( \frac{r}{r_c} \right)^{-\gamma} \exp \left[ -\left( \frac{r}{r_c} \right)^{2 - \gamma} \right],
\end{equation}
where ($\Sigma_0$) is a normalization constant, directly related to the disk gas mass ($M_{\mathrm{gas}}$), 
\begin{equation}
\Sigma_0 = (2 - \gamma) \frac{M_{\text{gas}}}{2 \pi {r_c}^2} \exp\left[ \left(\frac{r_{\text{in}}}{r_c}\right)^{2 - \gamma} \right].
\end{equation}
Given a specific temperature structure, the initial pressure at the disk midplane is calculated using the ideal gas law, and 
the vertical pressure profile is obtained by integrating the hydrostatic equilibrium equation. The density at each height 
is derived from the pressure using the ideal gas law and normalized to ensure that it integrates to the given surface density. 
This approach provides a self-consistent vertical density structure that reflects the combined effects of gravity, 
thermal pressure, and the specified temperature profile.

To explore the effect of the thickness of the VIRaM layer on the resulting CO isotopologue emission, 
we developed two disk models: one with a thick VIRaM layer and another with a thin one. 
Both models adopt the identical stellar and disk parameters listed in Table~\ref{tab:modelsetup}.
These parameters were chosen to approximate the well-studied Herbig~Ae system HD~163296 (see Section~\ref{sec:hd163296}).
The two models differ only in their prescribed vertical temperature structures: \\

\noindent
{\bf 1. Thin VIRaM Layer Model:}
The vertical temperature structure follows the commonly adopted analytic profile from \citet{Dartois+2003}, which approximates 
the physical models calculated by \citet{DAlessio+1999}, i.e.
   \[
   T(r, z) =
   \begin{cases} 
   T_{\text{atm}} + \left( T_{\text{mid}} - T_{\text{atm}} \right) \cos^3 \left( \frac{\pi z}{4H} \right) & \text{if } z < 2H, \\
   T_{\text{atm}} & \text{if } z \geq 2H,
   \end{cases}
   \]
 \noindent
where \(H\) is the gas scale height. In this model, the temperature increases rapidly from the midplane to the disk atmosphere, producing a thin VIRaM layer. \\

\noindent
{\bf 2. Thick VIRaM Layer Model:} 
The vertical temperature profile mimics the structure found by \citet{Qi+2011} from analyzing observations of the HD~163296 disk,
   \[
   T(r, z) =
   \begin{cases} 
   T_{\text{mid}} & \text{if } z < 2H, \\
   T_{\text{atm}} + \left( T_{\text{mid}} - T_{\text{atm}} \right) \cos^3 \left( \frac{\pi z}{12H} \right) & \text{if } 2H \leq z < 6H, \\
   T_{\text{atm}} & \text{if } z \geq 6H,
   \end{cases}
   \]
where the temperature is nearly isothermal up to 2\(H\) and transitions more gradually up to 6\(H\), 
creating an effectively thick VIRaM layer. \\

\begin{figure*}
\centering
\includegraphics[scale=0.8]{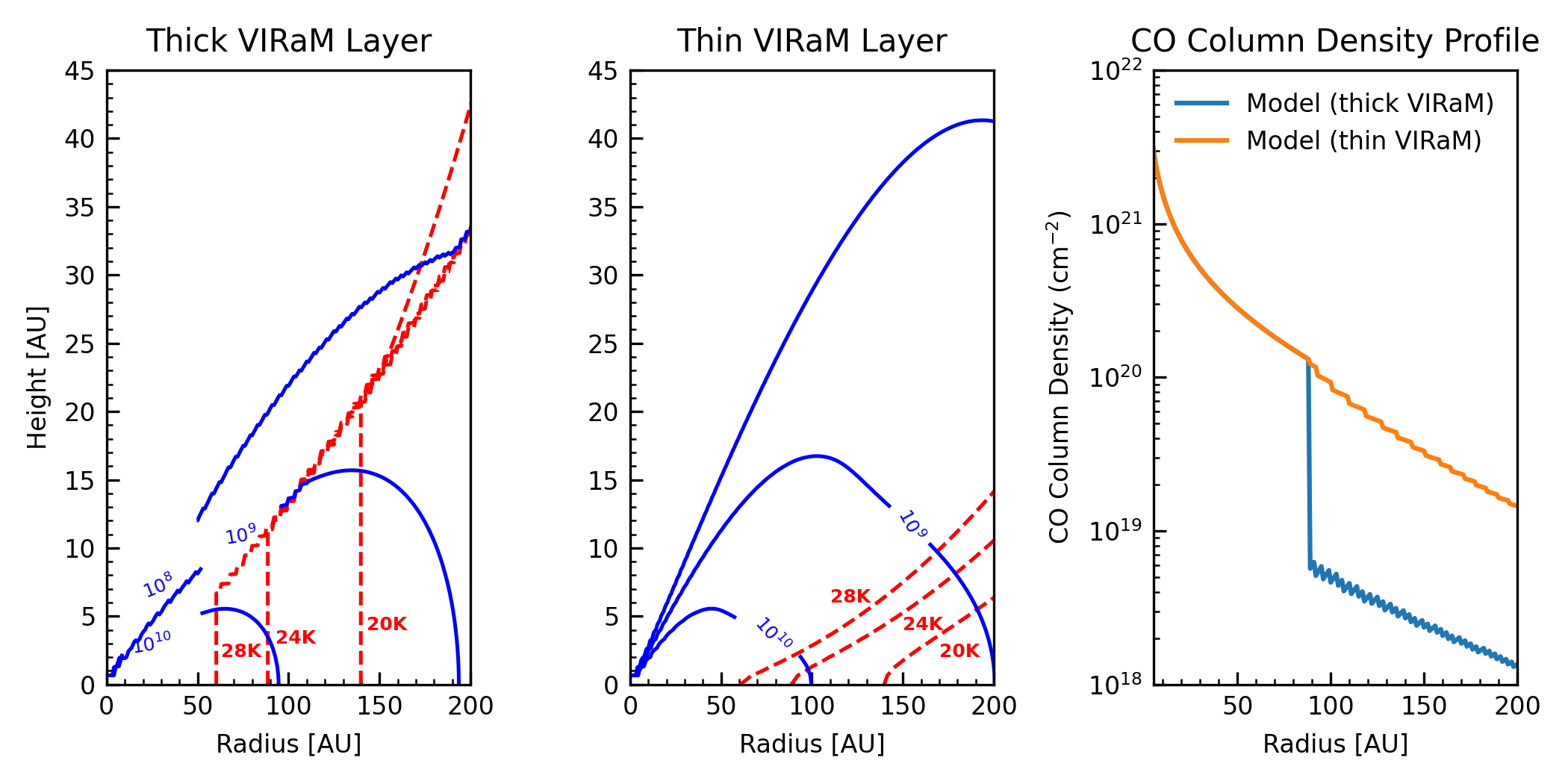}
\caption{Disk densities and temperatures for two models: (left) thick VIRaM layer with an isothermal vertical temperature structure, 
(center) thin VIRaM layer with a conventional vertical temperature gradient.
The temperature contours (red) are at 20, 24, and 28 K, and the density contours (blue) are at \(10^8\), \(10^9\), and \(10^{10}\) cm\(^{-3}\).
(right) CO column density distributions for the two models. The disk with a thick VIRaM layer shows a sharp radial drop across the CO snowline.
\label{fig:toymodels}}
\end{figure*}

For both models, the CO freeze-out temperature was assumed to be 24 K. The results are not sensitive to 
the exact choice of this parameter. For the thick VIRaM model, the corresponding CO snowline 
location is at 80~au. For the thin VIRaM layer model, which has all parameters the same other than the 
vertical temperature profile, the CO snowline is also 80 au. 
Adopting a distance of 100~pc, the snowline location lies at $0\farcs8$ radius. 
Figure~\ref{fig:toymodels} presents the temperature and density distributions for the two disk models. 
The left panel shows the model with the thick VIRaM layer model, with temperature contours nearly vertical up to 2\(H\). 
The middle panel shows the thin VIRaM layer model, which has a much steeper vertical temperature gradient
from the midplane to the atmosphere. 

\begin{table}[ht!]
\centering
\caption{Disk Model Parameters}
\label{tab:modelsetup}
\begin{tabular}{ll}
\hline
\textbf{Parameter}              & \textbf{Value}\\
\hline
Stellar Mass (\(M_*\))          & \(2 \, M_{\odot}\)\\
Disk Gas Mass (\(M_{\text{gas}}\))   & \(0.04 \, M_{\odot}\) \\
Inner Radius (\(r_{\text{in}}\)) & 1 au \\
Characteristic Radius (\(r_c\)) & 150 au \\
Surface Density Gradient (\(\gamma\)) & 0.9 \\
Midplane Temperature Profile (\(T_{\text{mid}}\)) & \(24 \left( \frac{R}{80\, \text{au}} \right)^{-0.4}\) \\
Atmosphere Temperature Profile (\(T_{\text{atm}}\)) & \(700 \left( \frac{R}{1\, \text{au}} \right)^{-0.55}\) \\
\hline
\end{tabular}
\end{table}

\noindent

We assume CO emission originates in a vertical layer of the disk with a constant abundance \( f_{\mathrm{CO}} = 1 \times 10^{-4} \). 
The upper boundary of this layer, toward the disk surface, is constrained by CO photodissociation, which occurs above an H$_2$ surface column density of approximately \( 10^{21} \, \text{cm}^{-2} \) \citep{Qi+2011}. The lower boundary, closer to the midplane, is defined by the CO freeze-out temperature, assumed to be 24~K. When the temperature drops below 24~K, the depletion factor -– the reduction in CO fractional abundance due to freeze-out -– is set to 320. In such regions, the CO abundance decreases to \( 3.1 \times 10^{-7} \). 
The right panel of Figure~\ref{fig:toymodels} displays the resulting CO column density profiles for both models. 
The model with a thick VIRaM layer exhibits a very sharp radial drop in CO column density at the snowline (80~au), 
whereas the model with a thin VIRaM layer shows a more gradual decrease.

These models serve to illustrate that the disk's vertical thermal structure significantly impacts the CO radial column density profile. 
For the thick VIRaM layer model, temperature contours in the disk region below the CO freeze-out temperature are nearly perpendicular to the 
density contours, which produces a very sharp drop in the CO column density across the snowline. In contrast, the thin VIRaM layer model shows 
a much smoother gradient in CO column density. 

In this study, we focus on the impact of the vertical thermal structure and CO depletion across the snowline, while intentionally omitting the effects of disk substructures, such as rings or gaps, which are not included in our current models. Future work will need to incorporate these substructures to assess how the small-scale temperature changes induced by such features \citep{ZhangS+2021} may affect the sharpness of the CO snowline transition. 

\subsection{CO Isotopologue Imaging Simulations}

The CO snowline can be probed best using optically thin tracers sensitive to cold gas, 
such as the lowest excitation rotational lines of rare CO isotopologues, together with sufficiently high spatial resolution -- in particular, 
a beam size smaller than half the CO snowline radius -- to resolve the column density change within the beam. 
Low optical depth provides intrinsic sensitivity to changes in the CO column density with the radius, while high spatial resolution prevents the 
shape of the radial profile from being smoothed out. 

Using the CO distributions from two models with thin and thick VIRaM layers and assuming local interstellar medium (ISM) ratios of 
CO/C\(^\mathrm{17}\)O = 2005 and CO/C\(^\mathrm{18}\)O = 557 \citep{Wilson1999}, 
we simulated \(\mathrm{C^{17}O}\) and \(\mathrm{C^{18}O}\) 1-0 and 2-1 emission using the nonlocal thermodynamic equilibrium, 
two-dimensional accelerated Monte Carlo radiative transfer code RATRAN \citep{Hogerheijde+2000}. 
The outputs of these simulations were then converted into synthetic visibility datasets 
using the \texttt{simobserve} task in the {\em Common Astronomy Software Application} package \citep{CASA+2022}. 

For the \(\mathrm{C^{17}O}\) and \(\mathrm{C^{18}O}\) 2-1 lines, we used ALMA for the imaging simulations. 
These simulations employed the configuration file \texttt{alma.cycle7.5.cfg}, 
which contains 43 antennas over baselines up to 2 km, resulting in a beam size of $0\farcs3$. 
This beam size is smaller than half the CO snowline radius ($0\farcs8$) in the models, 
providing sufficient resolution to detect the sharpness of the CO snowline transition in optically thin tracers. 
The simulated observations had an on-source integration time of 2 hours, 
aligning with the observations from the ALMA MAPS Large Program \citep{Oberg+2021}.

For the weaker \(\mathrm{C^{17}O}\) and \(\mathrm{C^{18}O}\) 1-0 lines, we used the ngVLA for the 
imaging simulations due to its superior sensitivity at these frequencies compared to ALMA. 
The ngVLA simulations used the configuration file \texttt{ngvla-revD.core.cfg}, 
which includes 114 antennas with a diameter of 18 meters within a 4 km region, 
providing a beam size of about $0\farcs3$, comparable to that for the ALMA simulations.
The simulated observations had an on-source integration time of 4 hours for ngVLA simulations. 
To add proper noise to the ngVLA simulations, we followed the recommendations from the 
CASA simulation guide\footnote{\url{https://casaguides.nrao.edu/index.php/Simulating_ngVLA_Data-CASA5.4.1}} 
and estimated the scaling parameter \texttt{$\sigma\_{simple}$}, then corrupted the simulated data using 
the \texttt{sm.setnoise} function of the \texttt{sm} toolkit.
Note that ALMA would require an integration time of about 30 times longer 
than ngVLA to reach the same noise level 
for these lines at this resolution,  according to the available online 
telescope sensitivity estimators
for ALMA\footnote{\url{https://almascience.nrao.edu/proposing/sensitivity-calculator}} (for typical band 3 weather)
and ngVLA\footnote{\url{https://ngect.nrao.edu/}}.

The detailed parameters for each simulated line are summarized in Table~\ref{tab:simulation_params}.
The source declination was assumed to be $-22$ degrees.
The disk was assumed to be inclined by $45$ degrees oriented at a position angle of $90$ degrees (east from north).
The simulations were conducted with 60 velocity channels of width 0.5~km~$^{-1}$ for each line. 
Imaging was performed using the \texttt{tclean} task in CASA with Briggs weighting and a robust parameter of $2$, 
i.e. close to natural weighting for maximum sensitivity. Finally, we collapsed the spectral cubes into integrated intensity maps using the Python package \code{bettermoments} \citep{Teague+2018}.

\begin{table*}
\caption{Imaging Simulation Parameters}
\label{tab:simulation_params}
\begin{tabular}{lcccccc}
\hline
{Line} & {Frequency} & {Channel} & {Int. Time}  & {Antenna}  & {Beam}  & {rms}  \\
~ & (GHz) & {(km~s$^{-1}$)} & {(s)} & {Config.} & {(arcsec)} & {(mJy~beam$^{-1}$)} \\
\hline
\(\mathrm{C^{17}O}\) 1-0 & 112.359 & 0.5 & 14,400 & \texttt{ngvla-revD.core.cfg} & $0.32 \times 0.20$ & 0.24 \\ 
\(\mathrm{C^{17}O}\) 2-1 & 224.714 & 0.5 & 7200 & \texttt{alma.cycle7.5.cfg}   & $0.33 \times 0.30$ & 1.03 \\   
\(\mathrm{C^{18}O}\) 1-0 & 109.782 & 0.5 & 14,400 & \texttt{ngvla-revD.core.cfg} & $0.32 \times 0.21$ & 0.18 \\ 
\(\mathrm{C^{18}O}\) 2-1 & 219.560 & 0.5 & 7200 & \texttt{alma.cycle7.5.cfg}   & $0.34 \times 0.31$ & 1.06 \\   
\hline
\end{tabular}
\end{table*}

\begin{figure*}
\centering
\includegraphics[width=0.49\textwidth]{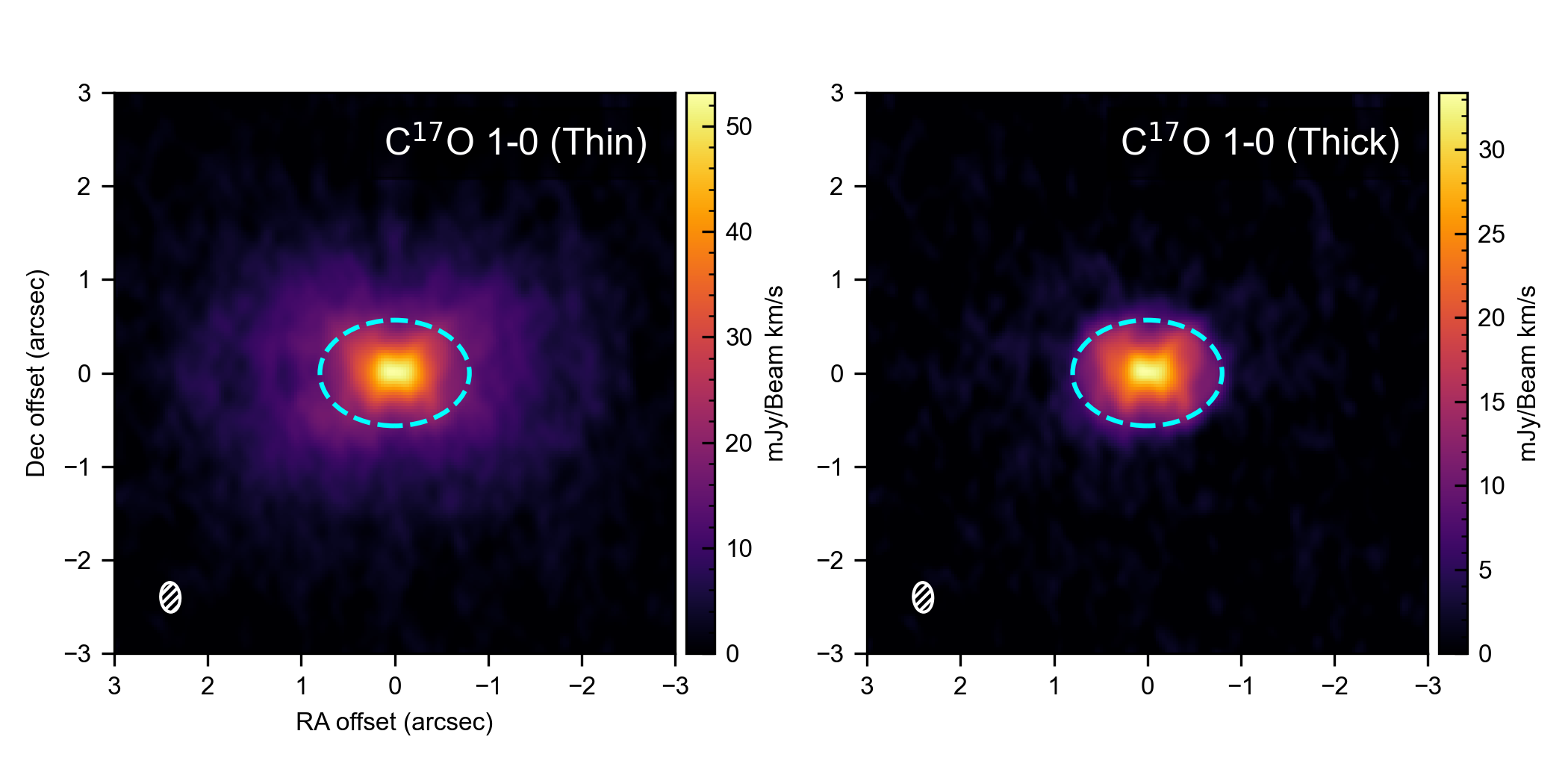} 
\includegraphics[width=0.49\textwidth]{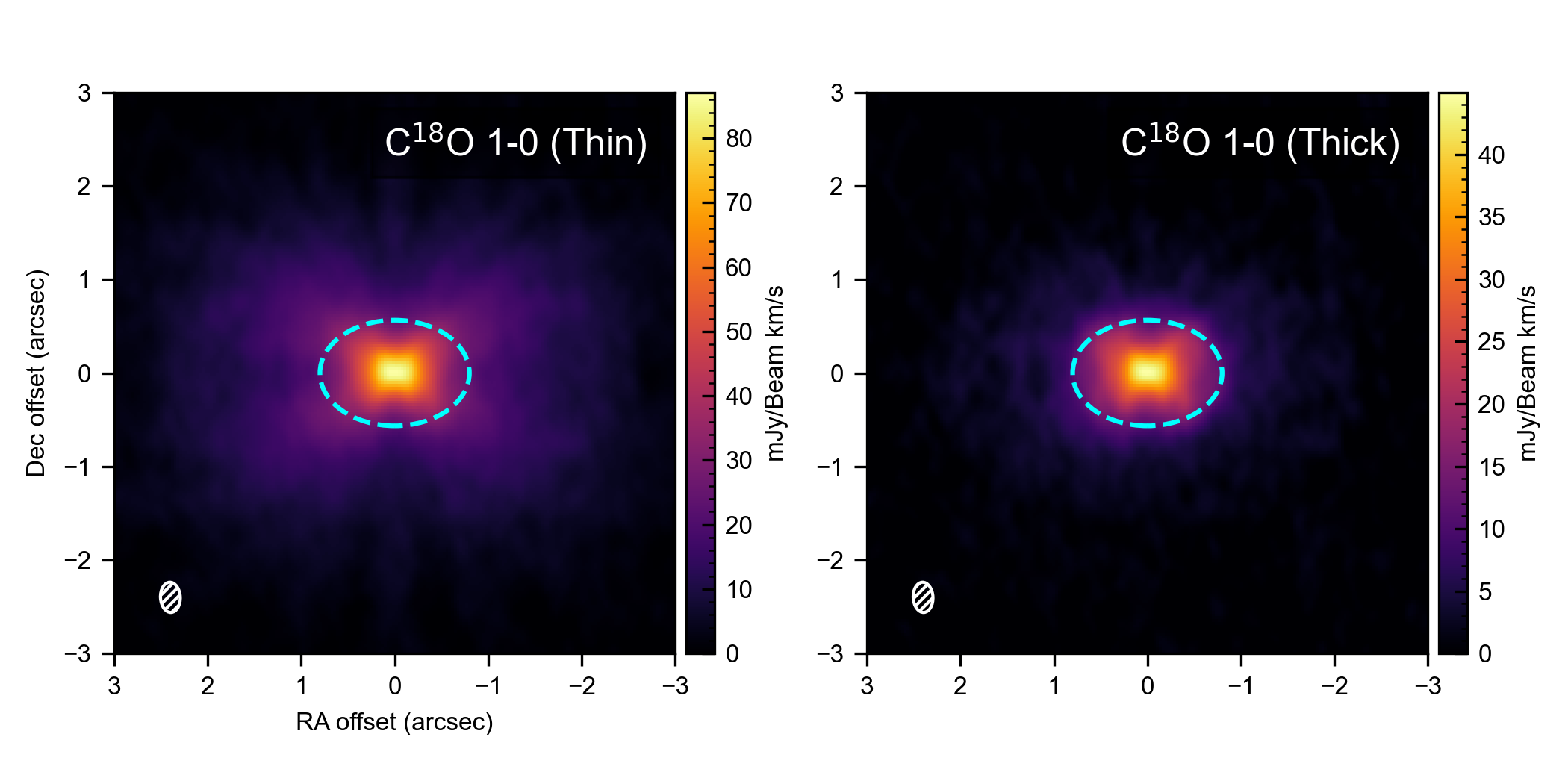} 
\includegraphics[width=0.49\textwidth]{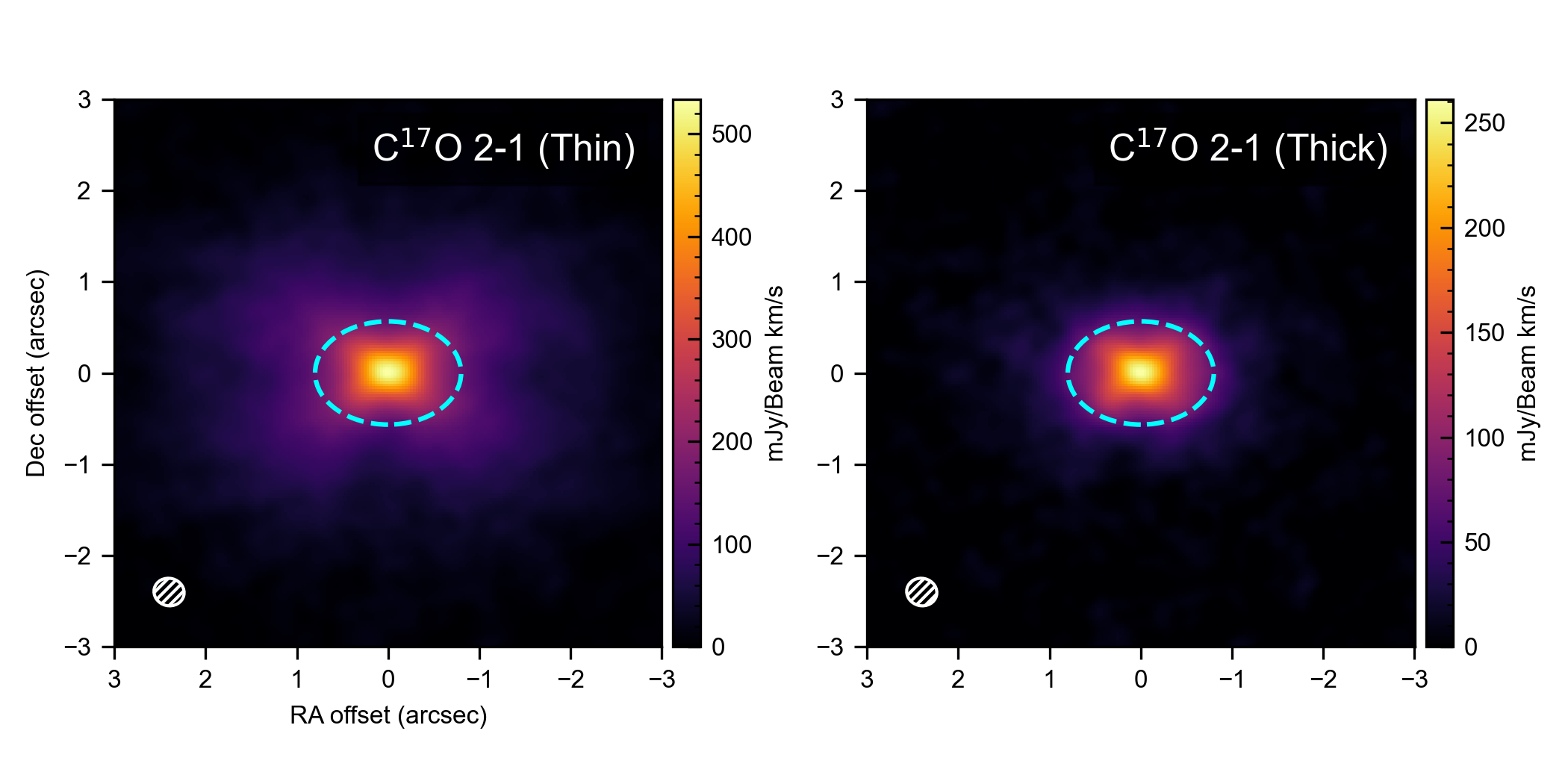}
\includegraphics[width=0.49\textwidth]{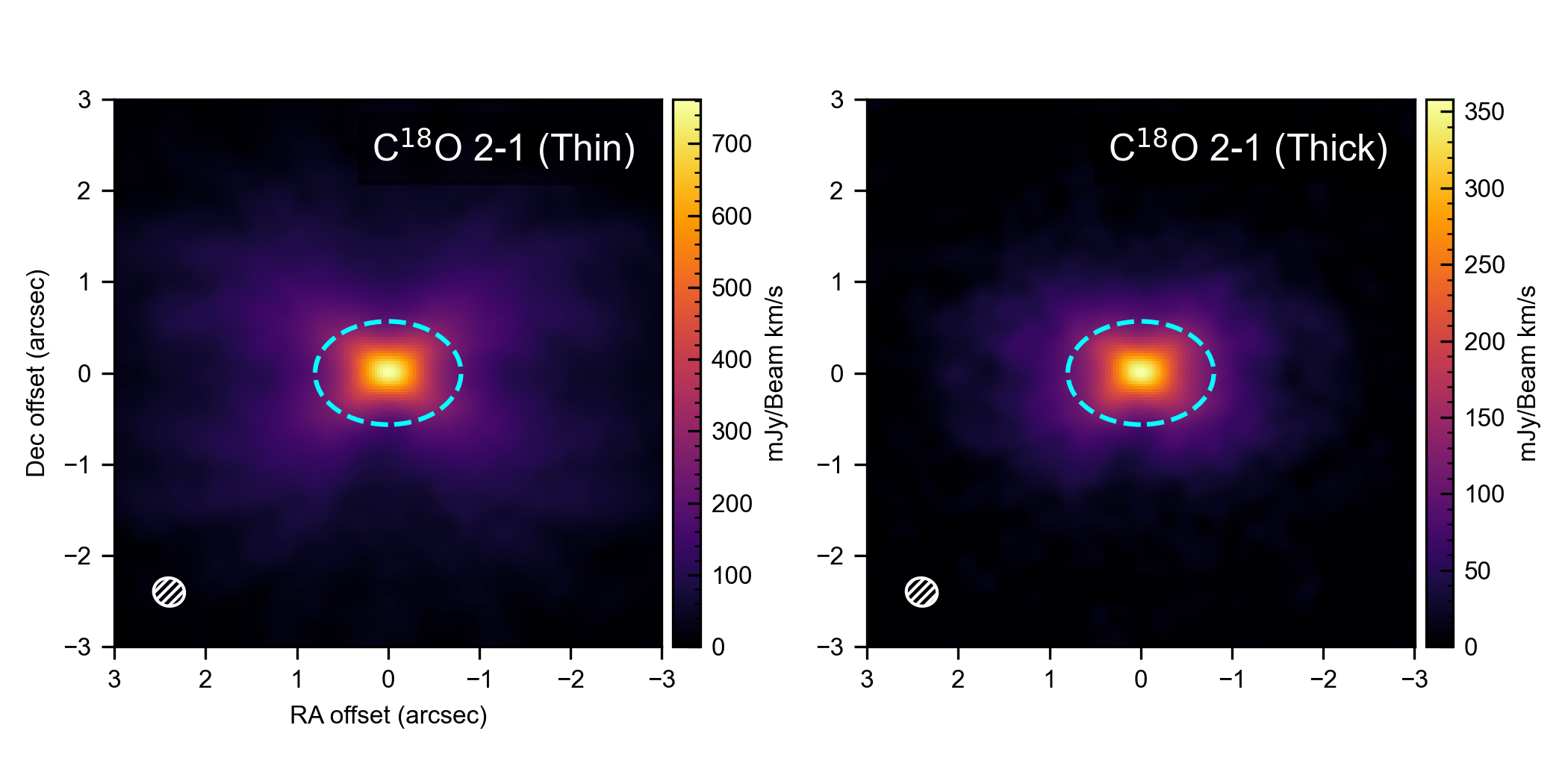}
\caption{Simulated integrated intensity images of C\(^{17}\)O 1-0 and C\(^{18}\)O 1-0 (ngVLA, top)  
and C\(^{17}\)O 2-1 and C\(^{18}\)O 2-1 (ALMA, bottom) for models with thin and thick VIRaM layers (as labeled).
The light blue dashed line indicates the location of the CO snowline in the models. 
These images reveal a much sharper drop in C\(^{17}\)O emission beyond the CO snowline for the thick VIRaM layer model
than for the thin VIRaM layer model. 
A similar, less pronounced, effect is visible in C\(^{18}\)O 1-0 but not C\(^{18}\)O 2-1 emission, as the
latter is optically thick.
\label{fig:toy1021-maps} }
\end{figure*}

Figure~\ref{fig:toy1021-maps} shows integrated intensity images of C\(^{17}\)O and C\(^{18}\)O 1-0 and 2-1 for the 
models with thin and thick VIRaM layers. The light blue dashed line indicates the location of the CO snowline. 
The radial extent of C\(^{17}\)O 1-0 emission appears smaller for the model with a thick VIRaM layer, 
as the intensity drops sharply beyond the CO snowline in this model. By contrast, the C\(^{17}\)O 1-0 emission extends 
much further beyond the CO snowline for the model with a thin VIRaM layer. 
Differences in emission extent are also apparent in the C\(^{17}\)O 2-1 and C\(^{18}\)O 1-0 images, albeit to a lesser 
extent in the latter as the optical depths of these lines are higher. Any differences in emission extent between
the two models are effectively obscured in the C\(^{18}\)O 2-1 images, as this line is optically thick for both. 

\begin{figure*}
\centering
\includegraphics[width=0.245\textwidth]{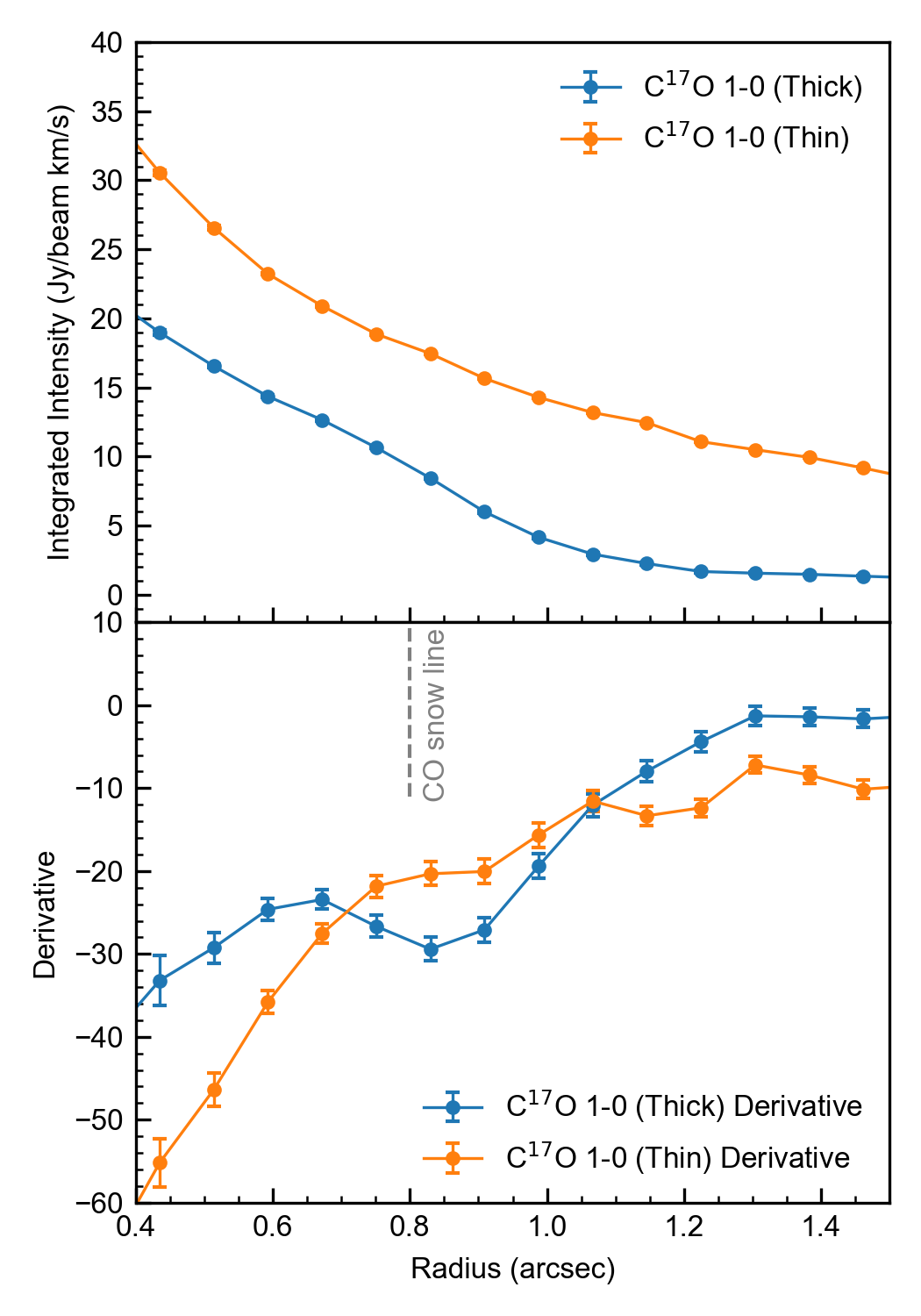}
\includegraphics[width=0.245\textwidth]{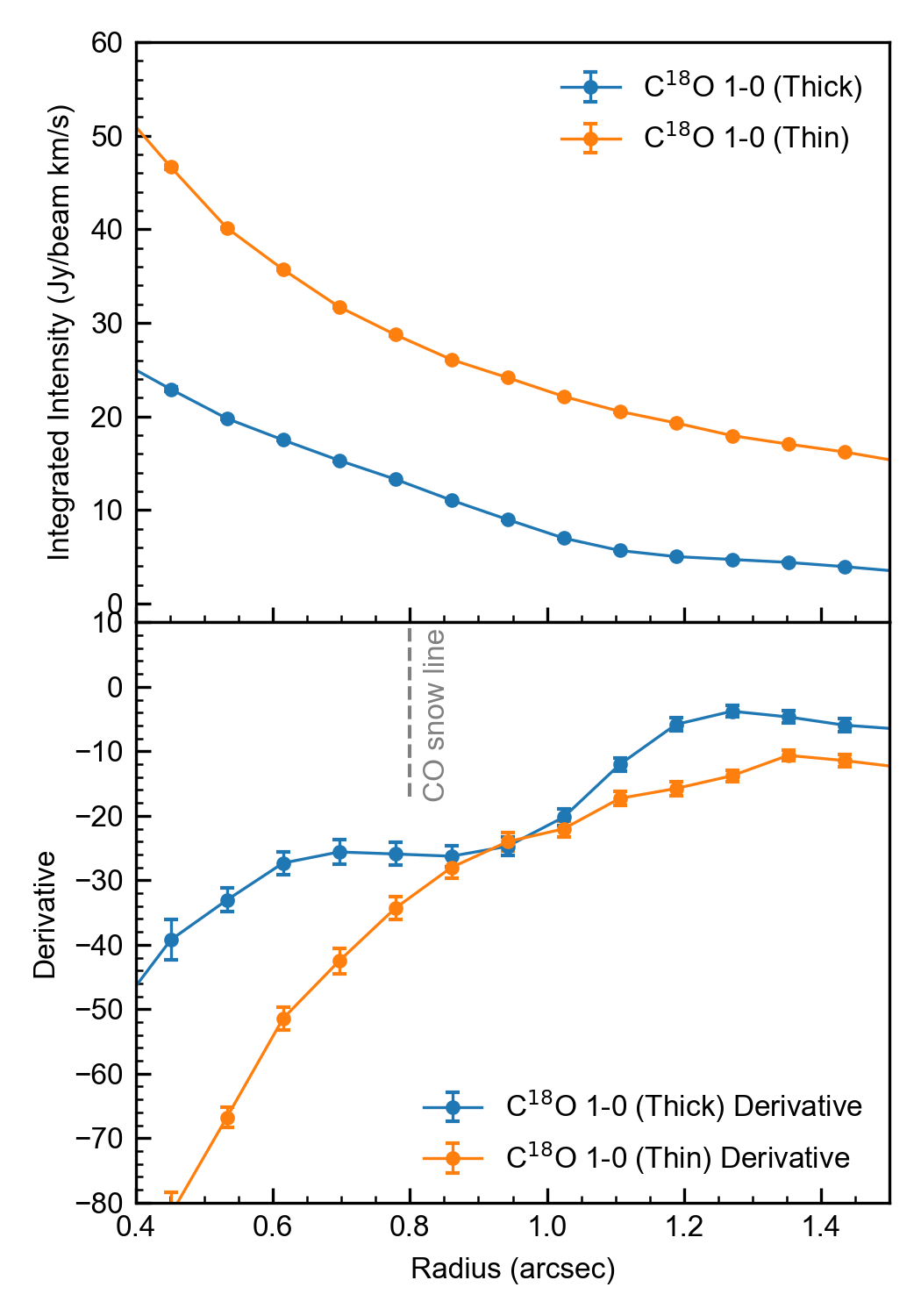}
\includegraphics[width=0.245\textwidth]{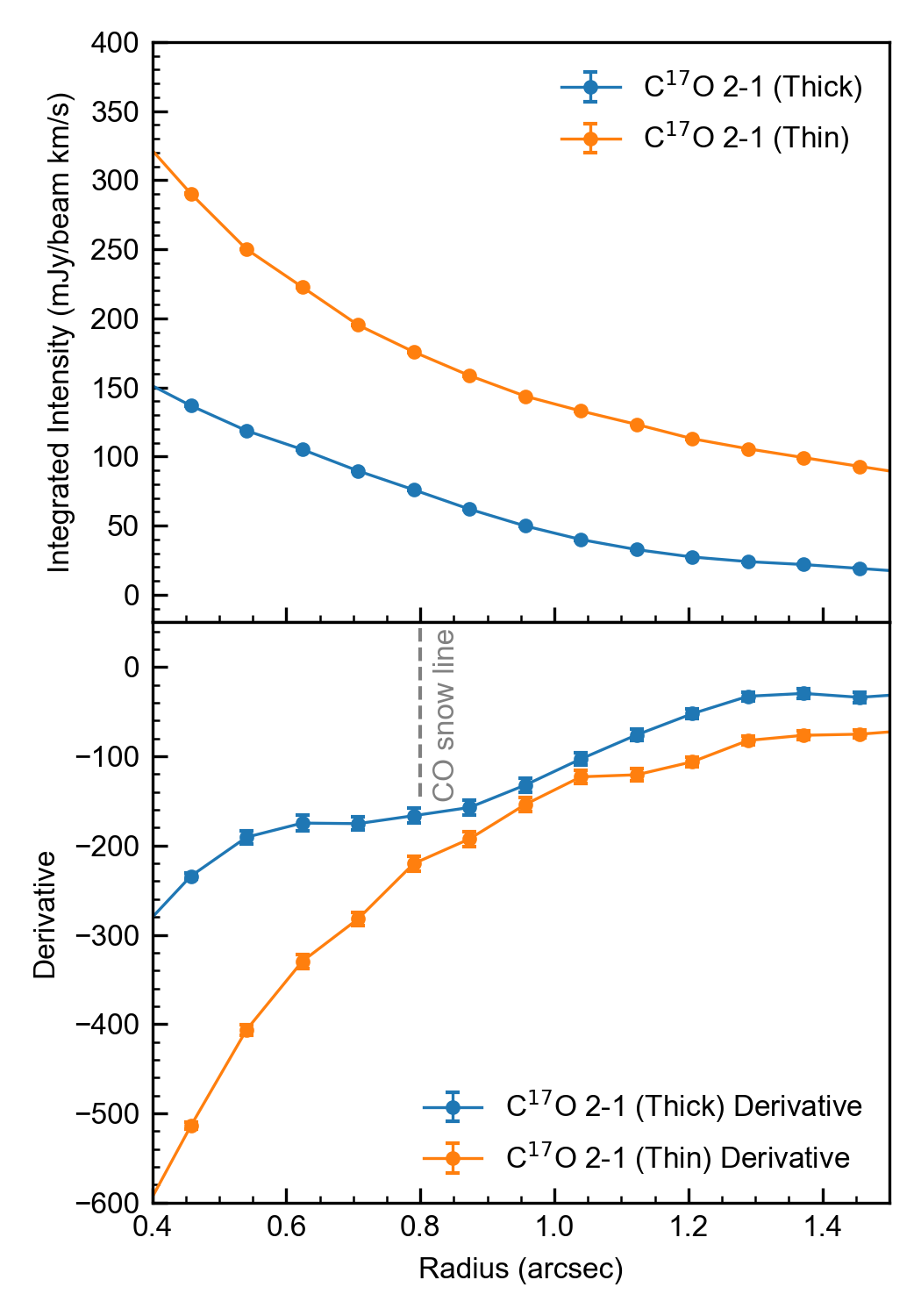}
\includegraphics[width=0.245\textwidth]{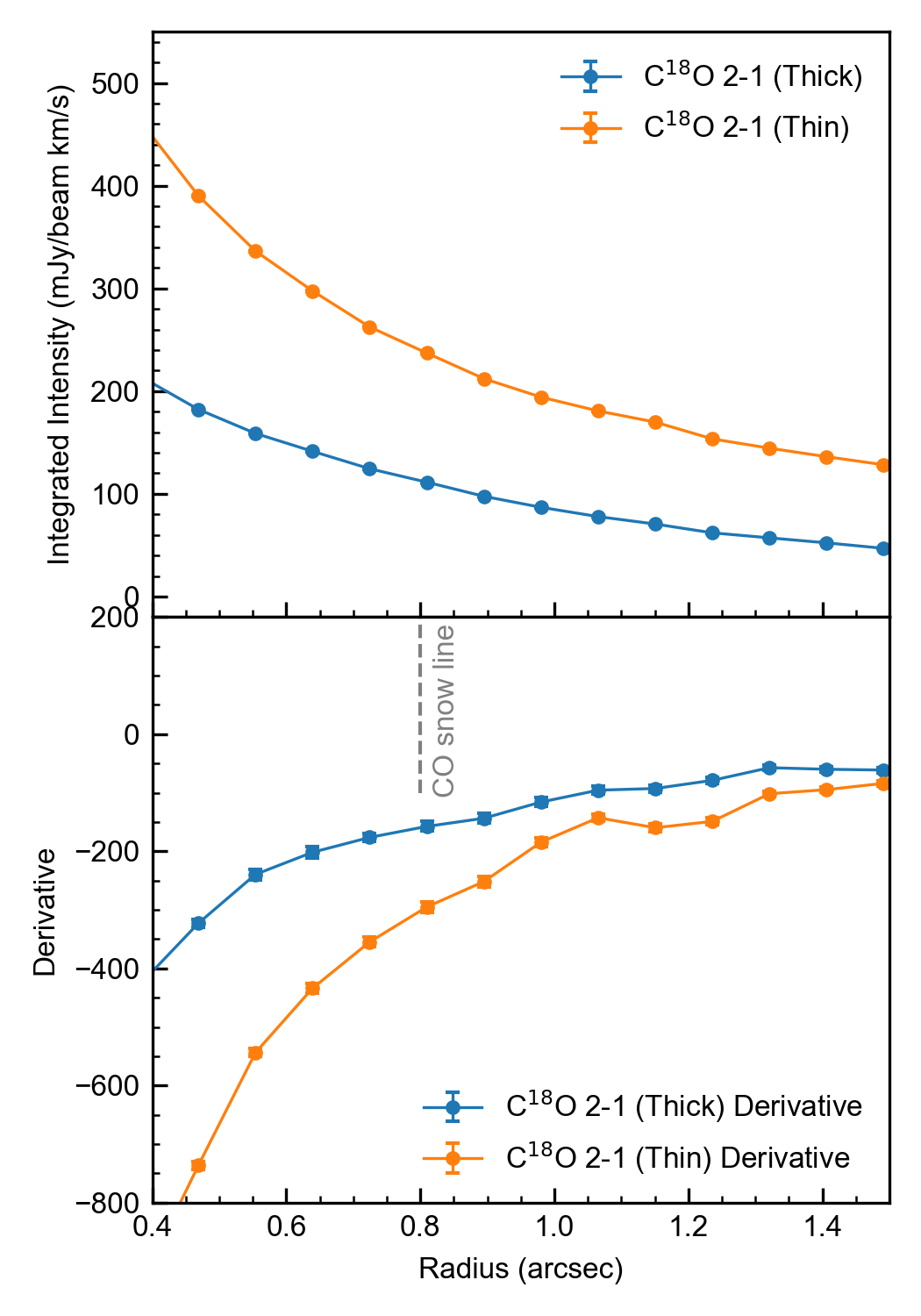}
\caption{Radial profiles of integrated intensity (top) and their derivatives (bottom) for simulated C\(^{17}\)O 1-0,  C\(^{18}\)O 1-0 
C\(^{17}\)O 2-1,  C\(^{18}\)O 2-1 emission (from left to right) in models with thick (blue) and thin (orange) VIRaM layers. 
The gray dashed line in each panel indicates the location of the CO snowline. 
For the optically thin lines, the derivative profiles for the thick VIRaM model show a minimum near the CO snowline resulting from
the significant change in the CO column density with the radius. The minimum in the derivative is less prominent in C\(^{18}\)O 1-0 
than in C\(^{17}\)O 1-0 due to increasing optical depth, and it becomes undetectable in C\(^{18}\)O 2-1, which is optically thick.
\label{fig:toy1021-profs} }
\end{figure*}

We extracted the deprojected and azimuthally averaged radial intensity profiles using the \texttt{radial\_profile} function 
from the Python package GoFish \citep{Teague2019}.
Figure~\ref{fig:toy1021-profs} (top panels) shows the radial profiles of the integrated intensity of the 
simulated C\(^{17}\)O and C\(^{18}\)O 1-0 and 2-1 emission for models with thick and thin VIRaM layers. 
While a sharp drop in the CO column density with the radius is not immediately evident from the radial profiles of 
the integrated emission intensity, it can be identified by analyzing the derivative of this radial profile. 
If the disk has a thick VIRaM layer, then the resulting sharp CO snowline will present a local minimum in the 
derivative, provided the emission is optically thin. 
Figure~\ref{fig:toy1021-profs} (bottom panels) shows the derivatives of the radial profiles, with the location 
of the CO snowline indicated by a gray dashed line. 

The derivatives for the thick VIRaM layer model show a clear minimum near the CO snowline for the J=1--0
lines, marking a significant change in the rate of CO column density decline with the radius. 
In contrast, the derivatives for the thin VIRaM layer model do not show a minimum 
clearly associated with the CO snowline. 
The minimum is less prominent in the C\(^{17}\)O 2-1 profile and is not visible at all in the 
C\(^{18}\)O 2-1 profile, as the emission from this more abundant isotopologue is optically thick throughout. 
This highlights that the ``sharpness'' of the CO snowline transition is best distinguished using the most 
optically thin tracers, particularly the ground-state lines. 
This radial profile derivative analysis provides an empirical way to determine the CO snowline location.

\section{Application to HD 163296}
\label{sec:hd163296}

The Herbig Ae star HD 163296 is surrounded by a disk that has been a favorite target of detailed study with the
improving capabilities of millimeter interferometry for more than 25 years 
\citep[e.g.,][]{Mannings+1997,Isella+2007,Qi+2011,Rosenfeld+2013,Oberg+2021}.
The N$_2$H$^+$ emission from this disk shows a characteristic bright, narrow ring with extended tenuous emission, 
which is well matched by the presence of a thick VIRaM layer \citep{Qi+2015} as found using the 
D'Alessio Irradiated Accretion Disk (DIAD) model \citep{Dalessio+2006} that provided 
a consistent fit to the spectral energy distribution and overall shape of resolved millimeter dust continuum
emission. This model features a refined vertical structure where dust settling is regulated by the transition location between small and large grains, 
$z=z_{big}$ in units of the gas scale height \(H\). 
This grain regulation enhances the temperature disparity between the disk surface and interior. Larger dust grains, with lower infrared opacities, 
are concentrated near the midplane, resulting in less efficient heating compared to smaller particles.
Consequently, the HD~163296 disk has a cold midplane with large grains at a significant scale height ($z_{big}=2H$). 
Near the midplane, the temperature remains approximately constant with height, effectively creating a thick VIRaM layer. 
Given the known CO snowline location around 0\farcs8 \citep{Qi+2015}, this disk serves as an ideal example to study 
the sharp drop in the CO column density across the CO snowline for a disk with a thick VIRaM layer.

\subsection{ALMA Observations of CO Isotopologues Toward HD~163296}
\begin{figure*}
\centering
\includegraphics[width=0.245\textwidth]{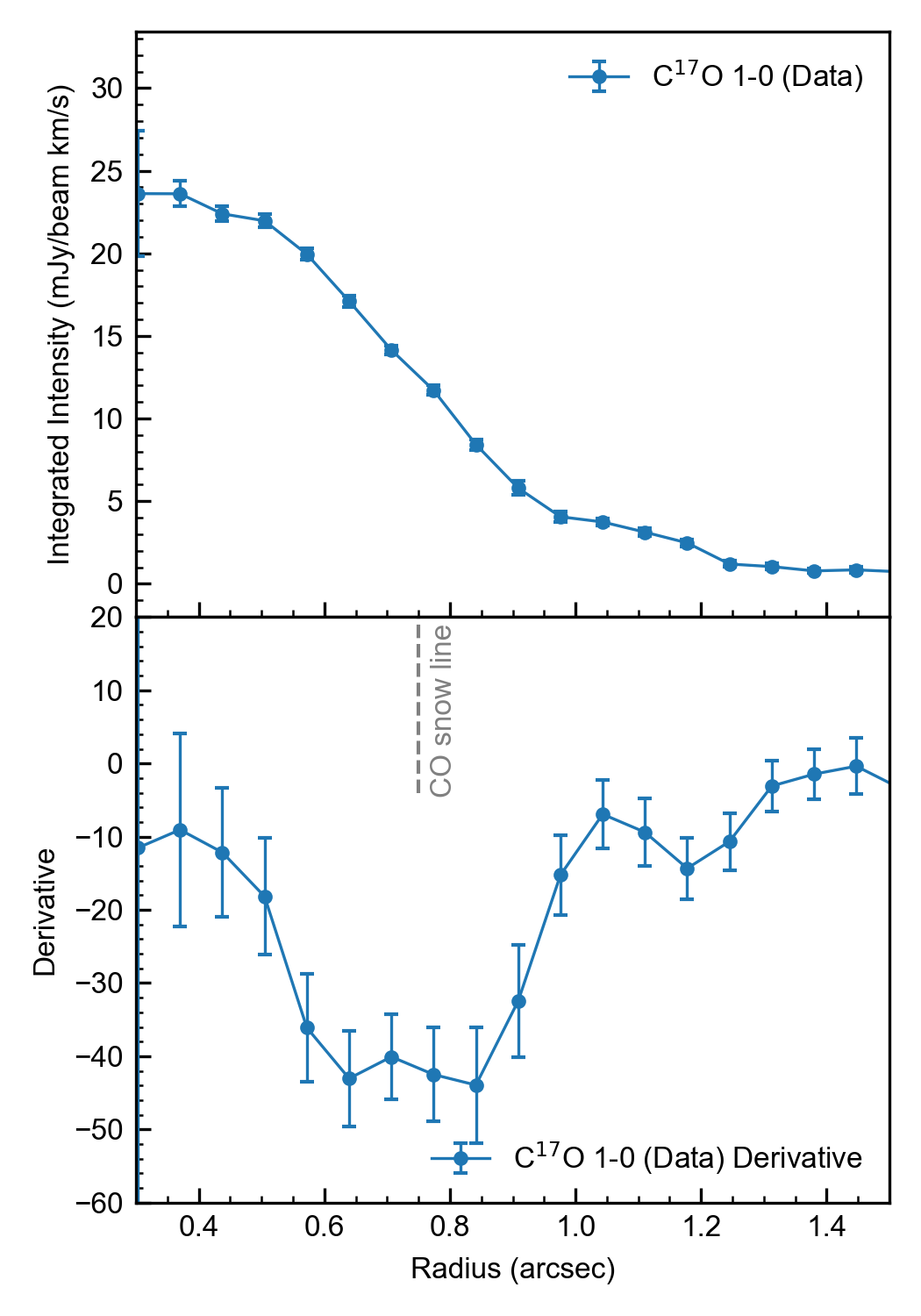}
\includegraphics[width=0.245\textwidth]{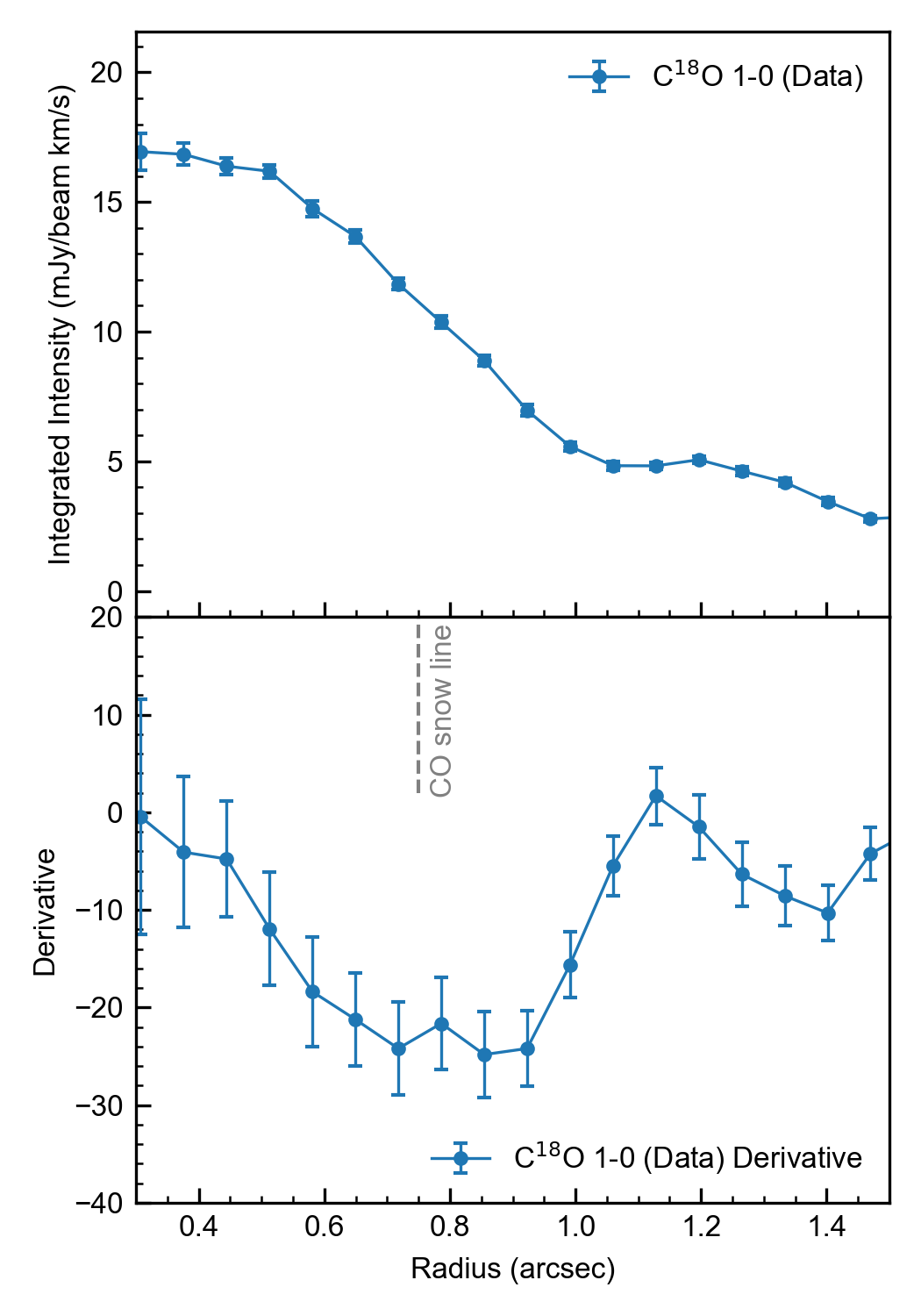}
\includegraphics[width=0.245\textwidth]{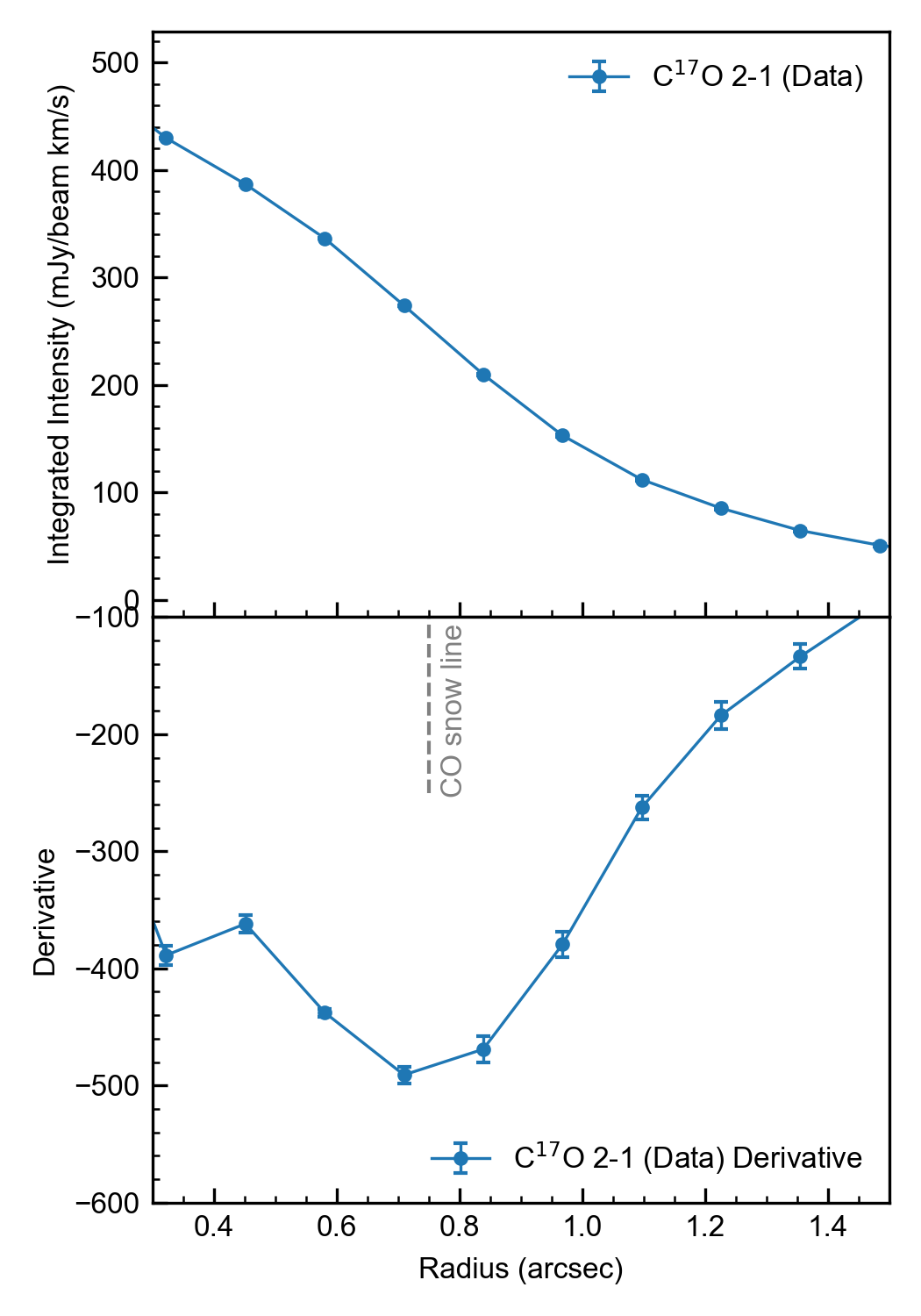}
\includegraphics[width=0.245\textwidth]{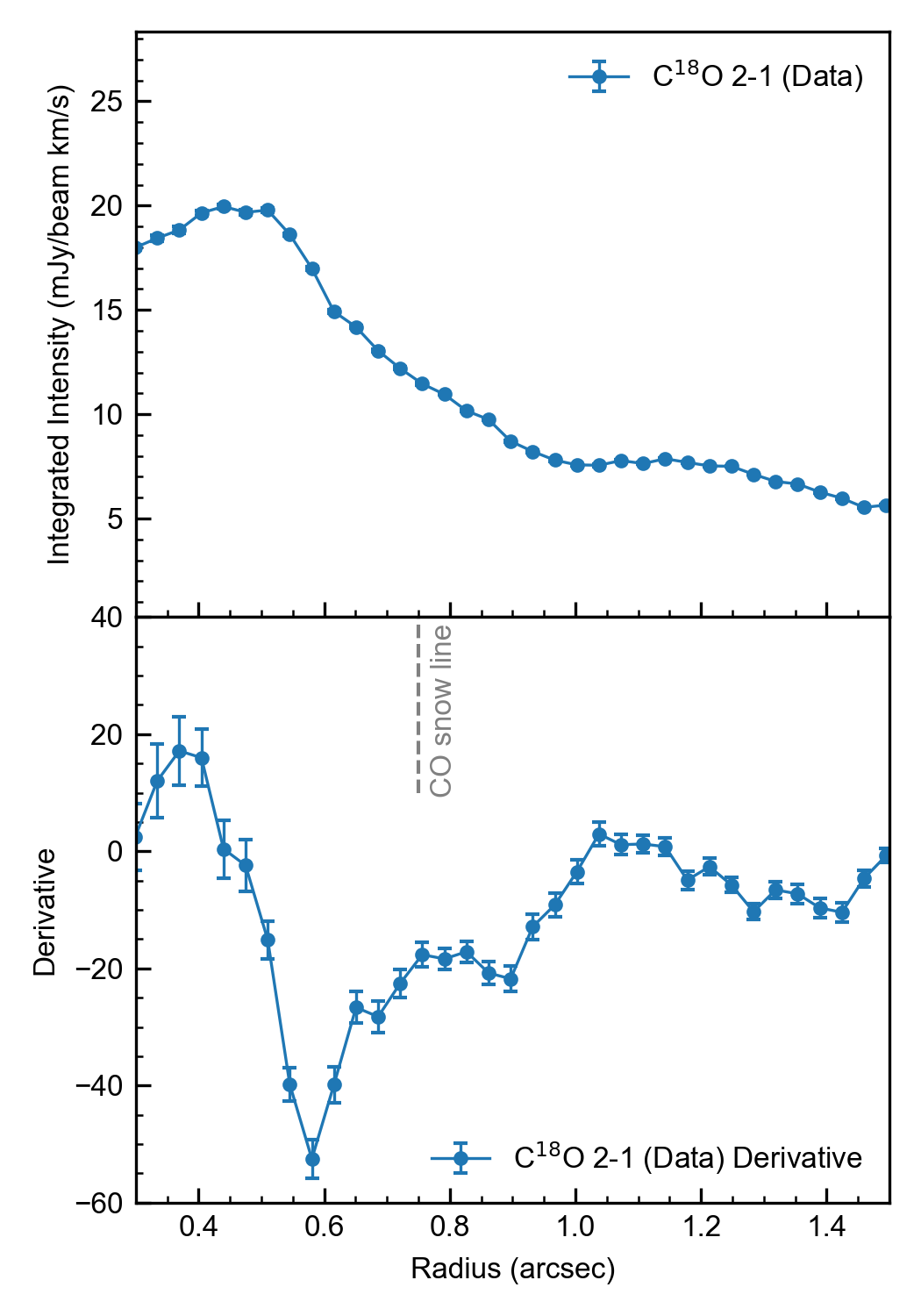}
\caption{Radial profiles and derivatives of \(\mathrm{C^{17}O}\) and \(\mathrm{C^{18}O}\) 1-0 and 2-1 emission observed by ALMA toward HD~163296. 
The gray dashed line in each panel indicates the location of the CO snowline. 
The local minima in the derivatives for the optically thin tracers align closely with the CO snowline location. 
\label{fig:hdrad}}
\end{figure*}

We examined archival ALMA observations of \(\mathrm{C^{17}O}\) and \(\mathrm{C^{18}O}\) 1-0 and \(\mathrm{C^{18}O}\) 2-1 lines in Bands 3 and 6 
toward the disk of HD~163296 from the MAPS Large Program (project 2018.1.01055.L). 
The observational details can be found in \citet{Oberg+2021}. 
The beam sizes for the \(\mathrm{C^{17}O}\) and \(\mathrm{C^{18}O}\) 1-0 observations are 0\farcs27\(\times\)0\farcs21, 
and for the \(\mathrm{C^{18}O}\) 2-1 observations, 0\farcs14\(\times\)0\farcs11. 
We also retrieved \(\mathrm{C^{17}O}\) 2-1 data from project 2016.1.00884.S (PI: V.V. Guzmán), with observational details provided in \citet{Carney+2019, Hernadez-vera+2024}. 
The beam size for the \(\mathrm{C^{17}O}\) 2-1 data is 0\farcs52\(\times\)0\farcs37.
Figure~\ref{fig:hdrad} shows the radial profiles of the integrated intensities and their derivatives. 
The derivatives of \(\mathrm{C^{17}O}\) 1-0, \(\mathrm{C^{18}O}\) 1-0, and \(\mathrm{C^{17}O}\) 2-1 all display a local minimum near the 
CO snowline location determined by \citet{Qi+2015}.  
The \(\mathrm{C^{18}O}\) 2-1 emission, being optically thick,  
is not suitable for this analysis, as expected from the models in Section~\ref{sec:toy}.

\subsection{\(\mathrm{C^{18}O}\) 1-0  in Other MAPS Disks}
\begin{figure*}[htbp]
\centering
\includegraphics[width=0.245\textwidth]{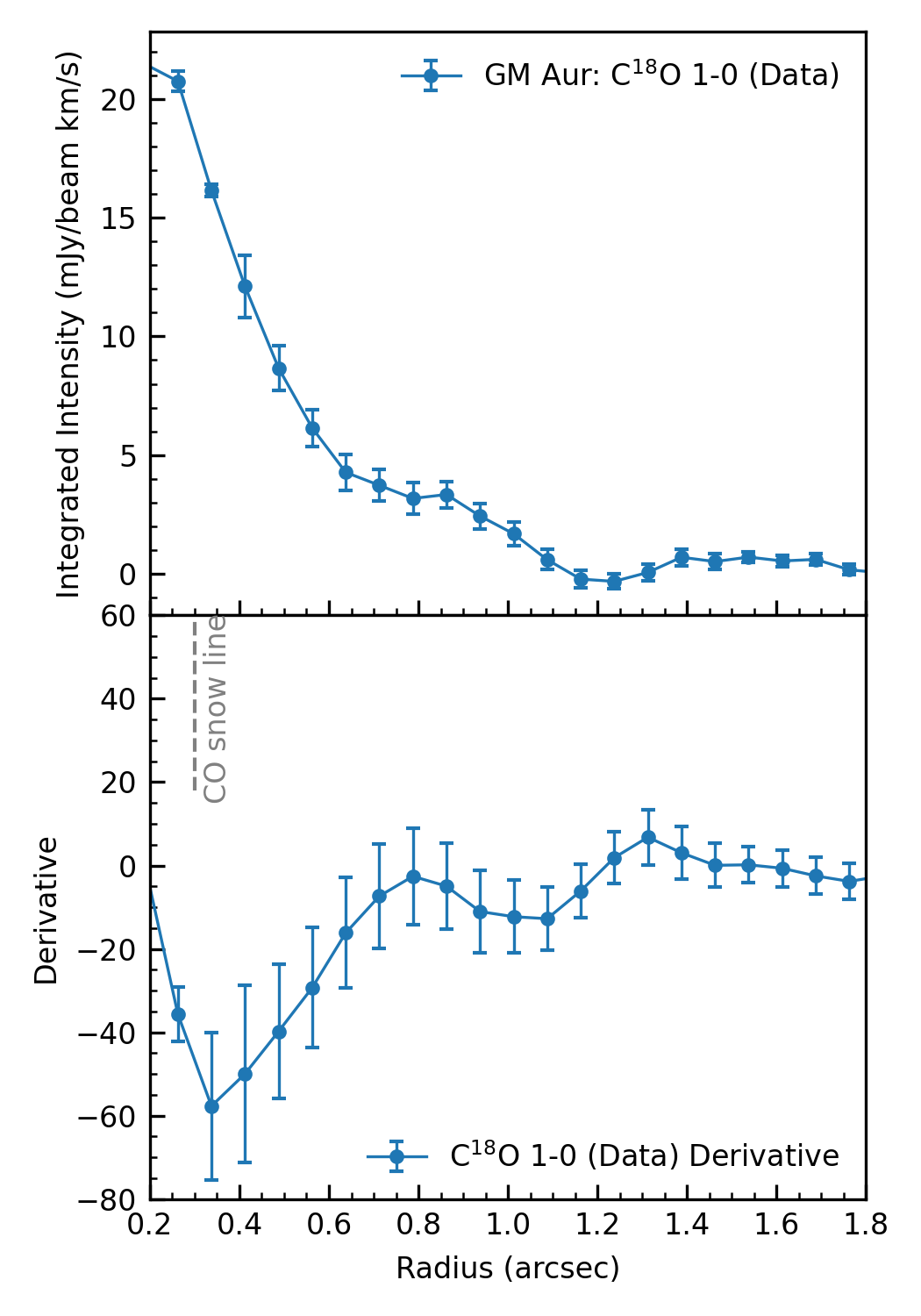}
\includegraphics[width=0.245\textwidth]{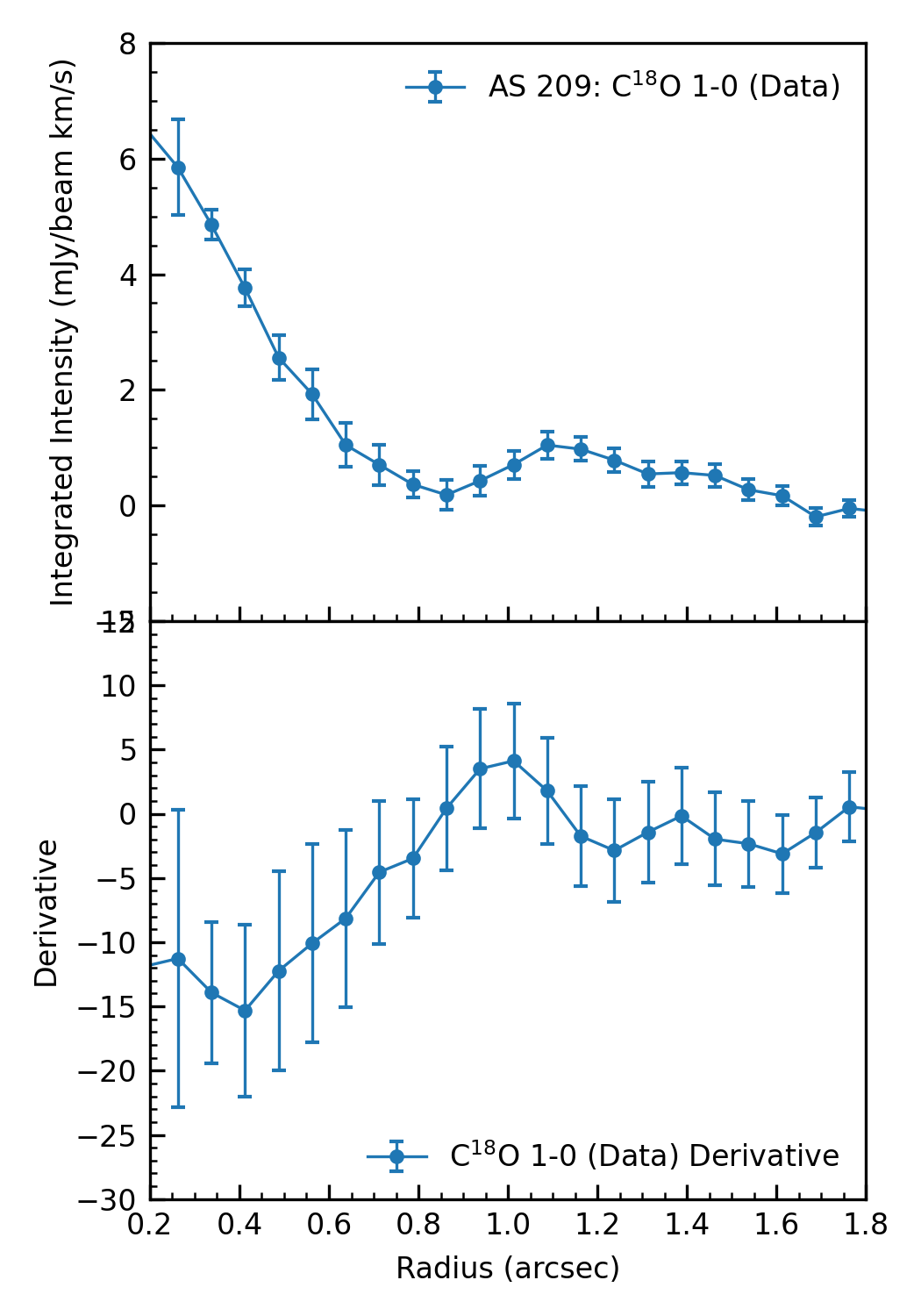}
\includegraphics[width=0.245\textwidth]{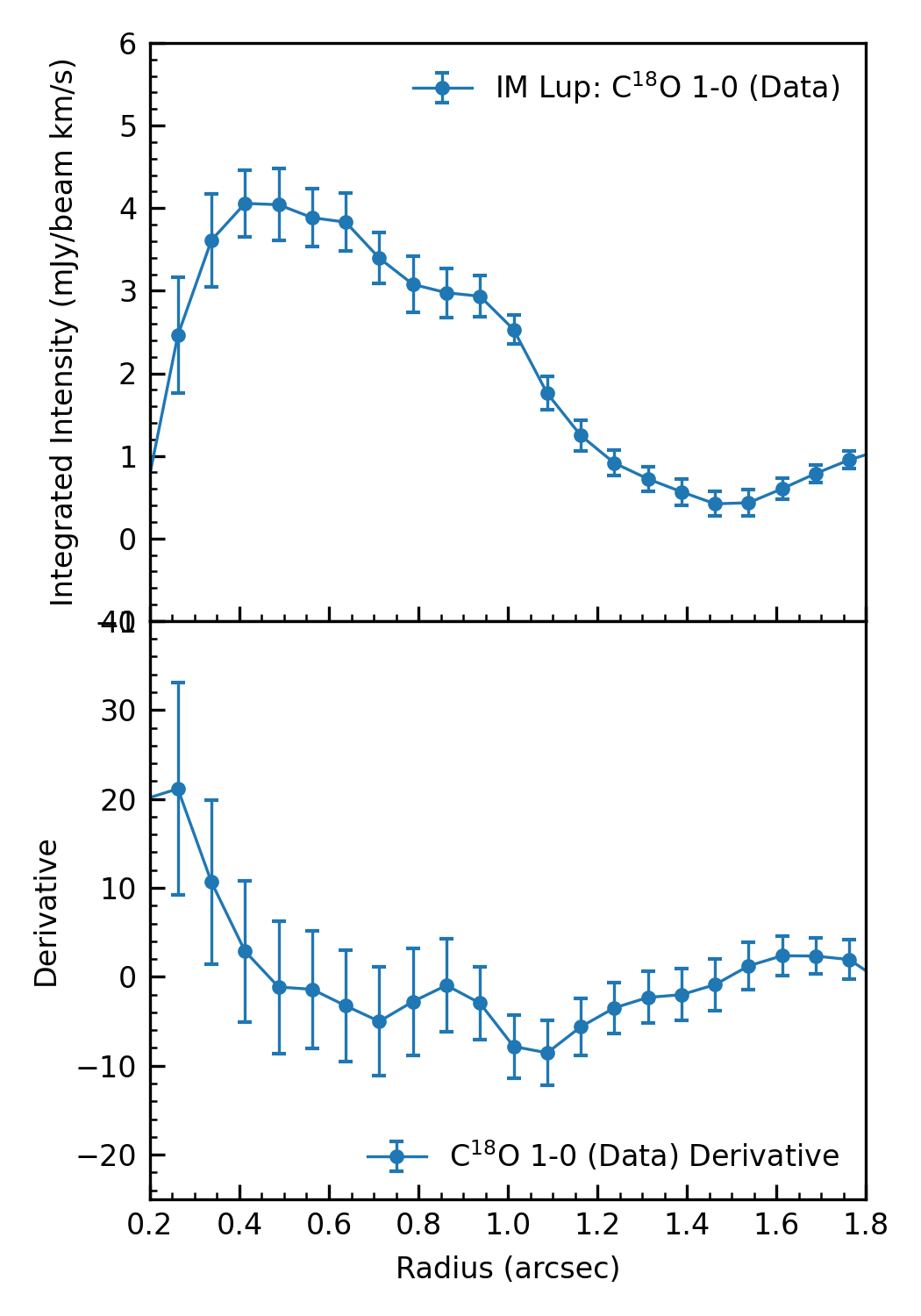}
\includegraphics[width=0.245\textwidth]{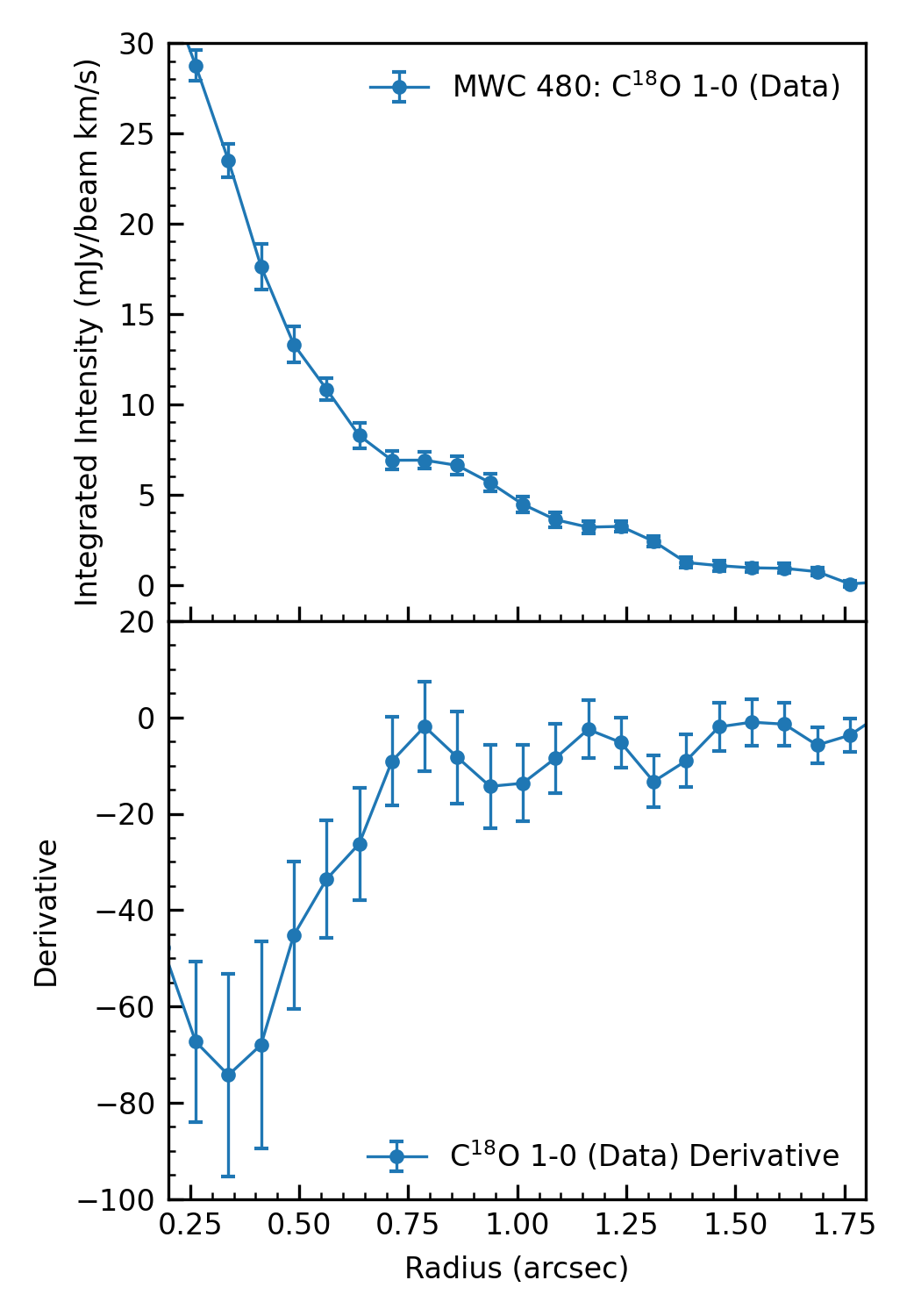}
\caption{Radial profiles and derivatives of \(\mathrm{C^{18}O}\) 1-0 emission observed by ALMA toward MAPS sources GM~Aur, AS~209, IM~Lup, and MWC~480. 
The gray dashed line in the panel of GM~Aur indicates the location of the CO snowline determined in Qi et al. (2019).
\label{fig:maps_radial}}
\end{figure*}

Since \(\mathrm{C^{18}O}\) is detected toward all of the disks in the MAPS program in addition to HD~163296, while \(\mathrm{C^{17}O}\) 1-0 is only marginal detected toward some of the MAPS disks,
we present the \(\mathrm{C^{18}O}\) radial profiles and derivatives in Figure~\ref{fig:maps_radial} for 
these disks -- GM~Aur, AS~209, IM~Lup, and MWC~480 -- with 0\farcs3 beam size 
(archival images from the MAPS Large Program project 2018.1.01055.L).

The GM Aur disk, identified by \citet{Qi+2019} as possessing a thick VIRaM layer, 
exhibits a well-defined CO snowline at 48 au, equivalent to 0\farcs3. As shown in Figure~\ref{fig:maps_radial} (GM Aur), 
despite the 0\farcs3 beam size being comparable to the snowline location, a clear dip in the derivative of the 
\(\mathrm{C^{18}O}\) 1-0 profile aligns with the expected snowline radius. This feature supports the hypothesis that 
a thick VIRaM layer facilitates a distinct, observable transition in the CO column density across the snowline.
Conversely, for the AS 209 and IM Lup disks, which were found to have thinner VIRaM layers, no evident dip 
can be seen in the \(\mathrm{C^{18}O}\) 1-0 derivative profiles. The absence of this feature could result from CO snowlines 
that are comparable to, or smaller than, the 0\farcs3 beam size, smoothing out the transition in the observations.

For the MWC 480 disk, an intriguing feature is observed: a dip in the derivative of the C\(^{18}\)O 1-0 radial profile at a radius smaller than 0\farcs3. 
However, this dip cannot be associated with the CO snowline, as thermal-chemical modeling of the MWC 480 disk places 
the CO snowline beyond 125 au (0\farcs8) \citep{Zhang+2021}. This discrepancy suggests that other physical processes, 
such as chemical evolution, substructures, or line opacity effects, similar to those that affect the C\(^{18}\)O 2-1 line
for HD 163296, are influencing the emission profile in this disk.

\subsection{CO Depletion Factor}
\label{sec:df}

To further investigate the effect of CO depletion across the snowline, we simulated the emission of CO isotopologues by varying the CO depletion factor ($df$). 
\citet{Qi+2015} derived a depletion factor of 5 ($df = 5$) to fit the ALMA \(\mathrm{C^{18}O}\) 2-1 data toward the HD~163296 disk, which was the most optically thin tracer available at that time. 
In this work, we assess whether a higher depletion factor could better explain the observed flux changes across the snowline, 
particularly for optically thinner tracers such as \(\mathrm{C^{17}O}\) and \(\mathrm{C^{18}O}\) 1-0. 
We followed the methodology of \citet{Qi+2015}, utilizing the RATRAN code to perform non-local thermodynamic equilibrium radiative transfer calculations 
and to generate synthetic CO data cubes. 
These synthetic data were sampled at the same spatial frequencies as the ALMA observations using the Python-based visibility sampling 
routine, \texttt{vis\_sample} \citep{Loomis+2018}. The deprojected and azimuthally averaged radial intensity profiles were then extracted using the \texttt{radial\_profile} function from GoFish.

\begin{figure*}
\centering
\includegraphics[scale=0.8]{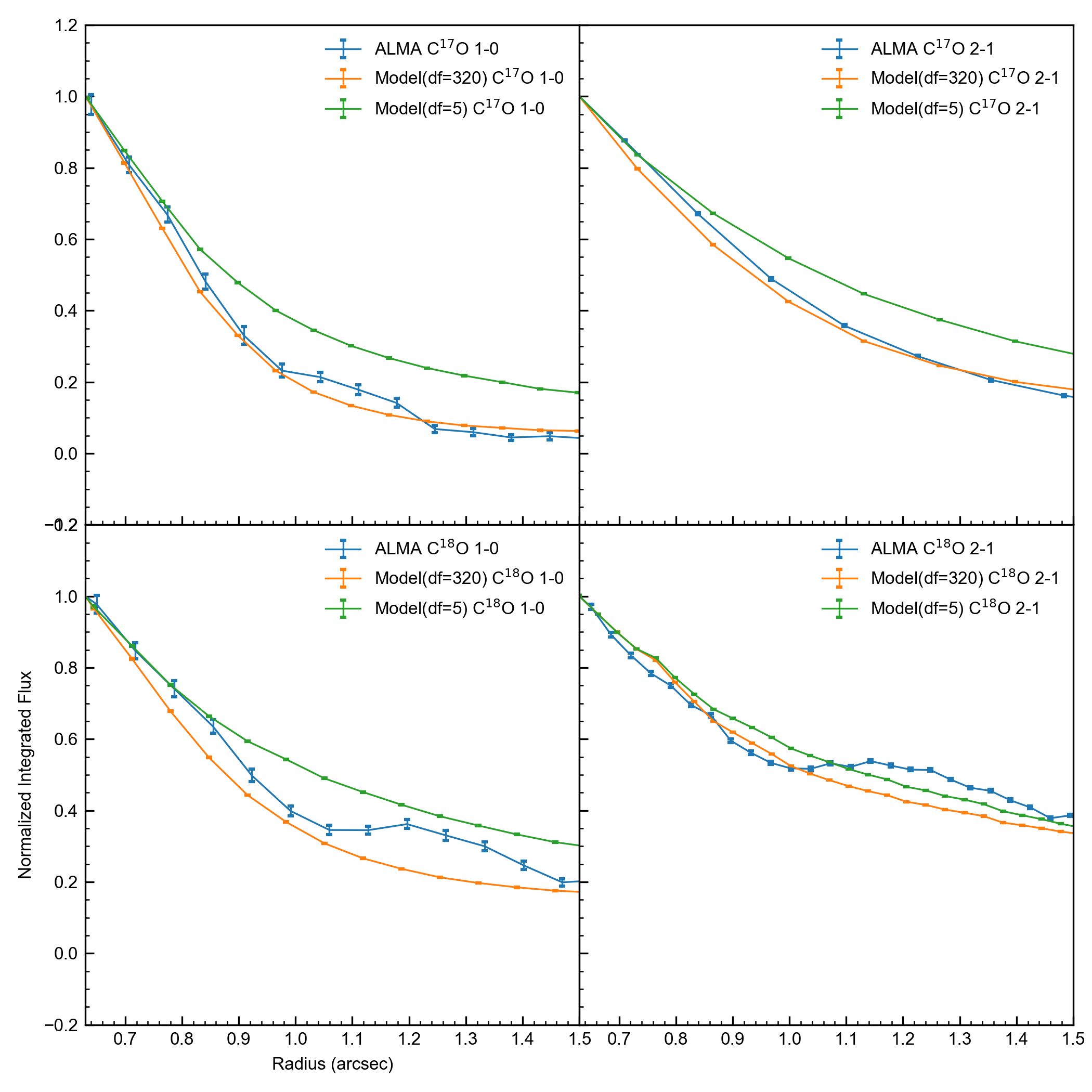}
\caption{Normalized flux profiles of HD~163296 \(\mathrm{C^{17}O}\) 1-0, 2-1 and \(\mathrm{C^{18}O}\) 1-0, 2-1 emissions for the ALMA observations (blue) 
and for disk models with CO depletion factors of 320 (orange) and 5 (green). The model with a depletion factor of 320 matches the data better. 
The \(\mathrm{C^{18}O}\) 2-1 emission is not sensitive to the CO depletion factor. 
\label{fig:hdnorm}}
\end{figure*}

While \citet{Qi+2015} fit the \(C^{18}O\) 2-1 data visibilities by varying both the CO freeze-out temperature (and, consequently, the location of the CO snowline) and the CO depletion factor (\( df \)), our study focuses on evaluating the local emission profile change across the CO snowline for the optically thinner CO isotopologues with various \( df \), avoiding the need to account for potential CO photodesorption by nonthermal processes beyond 200 au \citep{Salinas+2017}.
Since the CO snowline location has already been identified through the local minima in the derivatives of the emission profiles, as shown in Section~\ref{sec:hd163296}, we
explored models with depletion factors only, spanning a coarse logarithmic grid from 0.5 to 3.0 in steps of 0.5. To evaluate the relative flux change across the CO snowline, 
we normalized the flux profiles at a reference point near the snowline transition, specifically at 0\farcs65 (slightly interior to the snowline at 0\farcs8), 
to ensure sensitivity to the sharpness of the transition. 
Figure~\ref{fig:hdnorm} shows the normalized flux profiles of \(\mathrm{C^{17}O}\) and \(\mathrm{C^{18}O}\) 1-0 and 2-1 emissions for models with a 
depletion factor of \(df = 5\) (as derived by \citealp{Qi+2015}) and \(df = 10^{2.5}\) (i.e., 320), compared with the observations. The model with a 
depletion factor of 320 matches the data more closely than the model with a depletion factor of 5, particularly for the optically thinner tracers 
\(\mathrm{C^{17}O}\) 1-0, 2-1, and \(\mathrm{C^{18}O}\) 1-0. 
The similarity in the shapes of the \(\mathrm{C^{18}O}\) 2-1 flux profiles in both models and the observations 
indicates that this emission line is optically thick and largely insensitive to the depletion factor.

These findings from the ALMA observations of rare CO isotopologues are consistent with the presence of a thick VIRaM layer in the disk around HD~163296 and a sharp CO snowline. Such a structure significantly influences the CO isotopologue emission, highlighting the critical role of the disk's thermal structure.

\section{Discussion}
\label{sec:discussion}

We have shown that emission from rare CO isotopologues provides a useful
probe for the presence of a sharp CO snowline transition in disks with a
thick vertical isothermal layer. We provided direct evidence 
for this feature in the disk around the Herbig Ae star HD~163296 from 
low-$J$ CO isotopologue emission resolved by ALMA observations. 
We next discuss implications of this thermal structure for CO distribution 
models of HD~163296, the general use of CO isotopologue emission for estimating 
total disk gas masses, and the feasibility of future observations in more disks. 

\subsection{Comparison of HD 163296 CO Distribution Models}

\begin{figure*}
\centering
\includegraphics[scale=1]{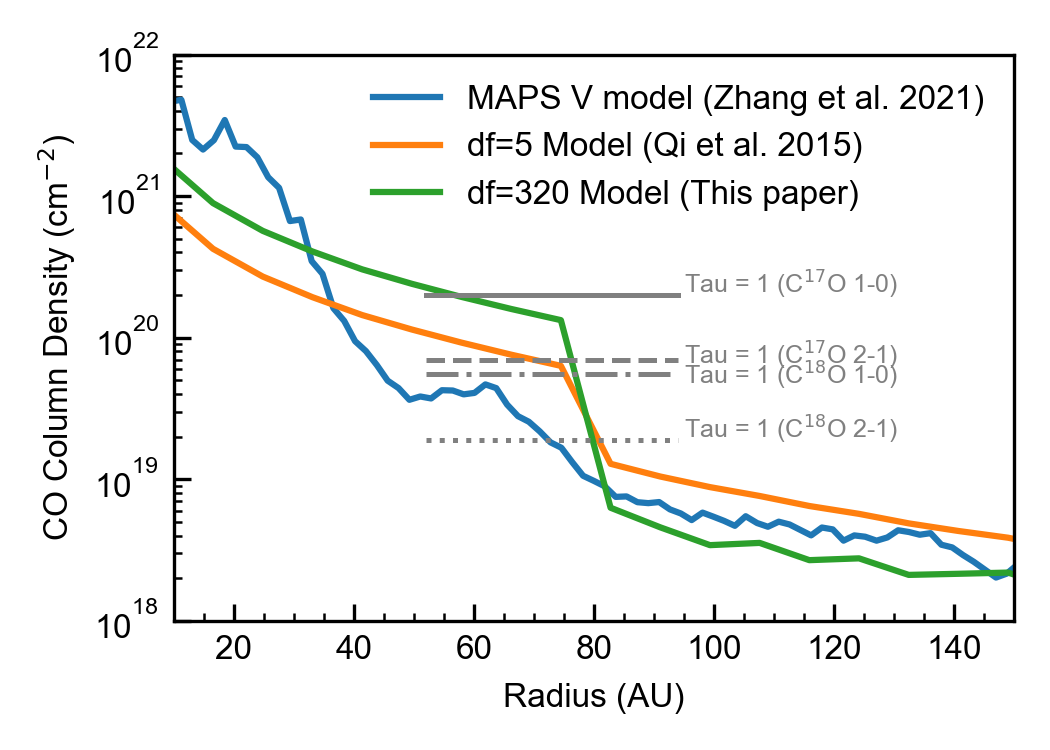}
\includegraphics[scale=1]{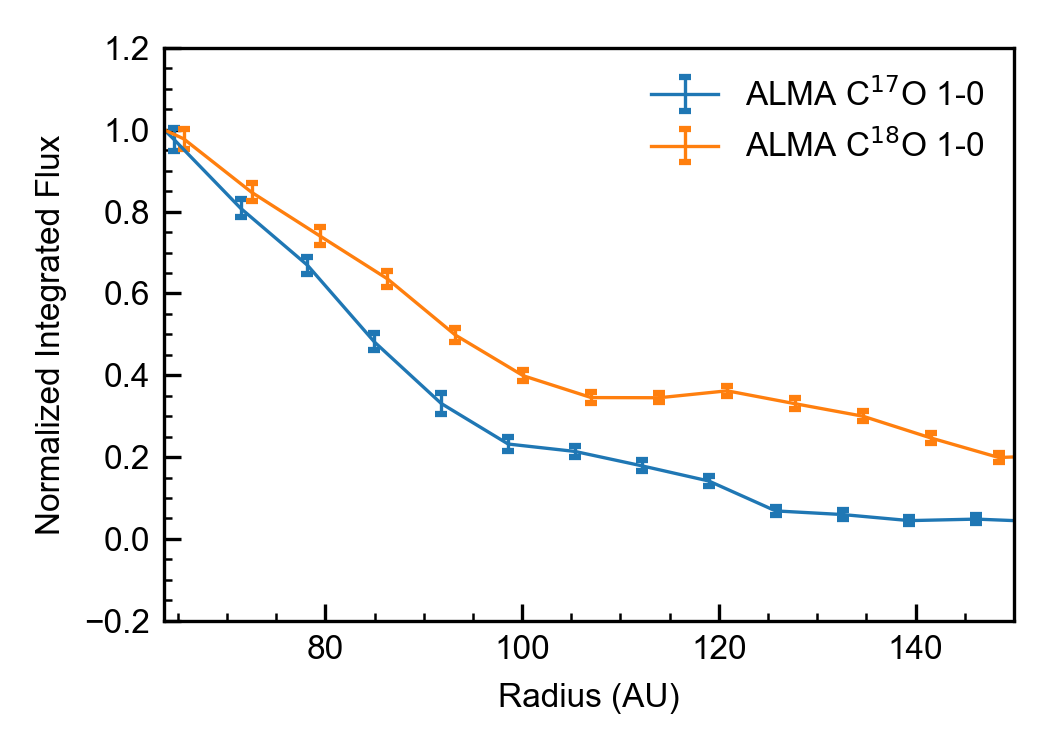}
\caption{
{\it Left}: comparison of CO column density radial profiles for three different models of the HD~163296 disk:
MAPS model from \citet{Zhang+2021} (blue), model with a CO depletion factor of 5 from 
\citet{Qi+2015} (orange), and model presented in this paper with a CO depletion factor of 320 (green). 
Reference CO column density values corresponding to \(\tau = 1\) for different CO isotopologue transitions 
based on RADEX calculations (see text) are marked in gray: 
\(2 \times 10^{20} \, \text{cm}^{-2}\) for C\(^{17}\)O 1-0, 
\(7 \times 10^{19} \, \text{cm}^{-2}\) for C\(^{17}\)O 2-1, 
\(5.5 \times 10^{19} \, \text{cm}^{-2}\) for C\(^{18}\)O 1-0, and 
\(1.9 \times 10^{19} \, \text{cm}^{-2}\) for C\(^{18}\)O 2-1,
assuming ISM ratios of CO/C\(^{17}\)O = 2005 and CO/C\(^{18}\)O = 557. 
{\it Right:} radial profiles of the normalized integrated flux of C\(^{17}\)O 1-0 (blue) and 
C\(^{18}\)O 1-0 (orange) emission observed with ALMA toward the disk of HD~163296. 
The profiles are normalized to their respective values at a radius of 70 au to facilitate comparison of 
the relative flux changes. Error bars indicate the observational uncertainties. 
The relative flux of C\(^{18}\)O 1-0 remains consistently higher than that of C\(^{17}\)O 1-0, 
suggesting that C\(^{18}\)O 1-0 is optically thicker than C\(^{17}\)O 1-0 across the CO snowline. 
\label{fig:hdcol}}
\end{figure*}

The sensitivity of the low-$J$ CO isotopologue lines to the CO snowline structure is underscored by comparing the emission resulting from different disk models. Figure~\ref{fig:hdcol} (left panel) shows the radial profile of the CO column density for three models of the disk around HD~163296: (1) the model with a depletion factor (\(df\)) of 5, as derived by \citet{Qi+2015}; (2) the model with a higher \(df\) of 320, which provides a better match to the data, as demonstrated in Section~\ref{sec:df}; and (3) the model from the ALMA MAPS program, where the analysis of CO isotopologues suggested an enhancement in CO abundance interior to the snowline \citep{Zhang+2021}.
The key region of interest in this comparison lies between 70 and 90 au, where the CO column density shows significant variations across the CO snowline. The model with \(df = 320\) demonstrates a much steeper decline in the CO column density compared to either the \(df = 5\) model or the MAPS model. In particular, the CO column density in the \(df = 320\) model decreases by more than a factor of 20 between 70 and 90 au, while the \(df = 5\) model shows a decrease by a factor of 5, and the MAPS model shows a decrease by only a factor of 2. This pronounced drop in the \(df = 320\) model provides clear evidence of a sharp CO snowline transition in the disk of HD~163296.

Figure~\ref{fig:hdcol} (left panel) also indicates reference CO column density values corresponding to optical depths 
(\(\tau\)) of 1 for the different CO isotopologue lines, as calculated using the RADEX code \citep{vanderTak+2007} and assuming 
interstellar medium ratios of CO/C\(^{17}\)O = 2005 and CO/C\(^{18}\)O = 557. 
The grey lines indicate the column densities at which the emission becomes optically thick: 
\(2 \times 10^{20} \, \text{cm}^{-2}\) for \(\tau = 1\) of C\(^{17}\)O 1-0, 
\(7 \times 10^{19} \, \text{cm}^{-2}\) for \(\tau = 1\) of C\(^{17}\)O 2-1, 
\(5.5 \times 10^{19} \, \text{cm}^{-2}\) for \(\tau = 1\) of C\(^{18}\)O 1-0, and 
\(1.9 \times 10^{19} \, \text{cm}^{-2}\) for \(\tau = 1\) of C\(^{18}\)O 2-1.
From 70 to 90 au, all models predict CO column densities below the threshold for \(\tau = 1\) of 
C\(^{17}\)O 1-0, indicating that this transition remains optically thin across the CO snowline.
However, both the $df = 320$ and the $df = 5$ models predict CO column densities above the threshold 
for \(\tau = 1\) of C\(^{18}\)O 1-0, indicating that this transition is optically thick; 
the MAPS model remains below this threshold, implying that C\(^{18}\)O 1-0 is optically thin.

Figure~\ref{fig:hdcol} (right panel) shows the radial profile of the integrated flux of 
C\(^{18}\)O 1-0 compared to C\(^{17}\)O 1-0, normalized at 70~au to more clearly show the difference 
in profile shape. The steeper decline of emission from C\(^{17}\)O compared to C\(^{18}\)O aligns 
with the predictions of the models with $df=5$ and $df=320$ where C\(^{18}\)O 1-0 emission is 
indeed optically thicker than the C\(^{17}\)O 1-0 emission across the CO snowline, 
but contradicts the expectations of the MAPS model where both lines would remain optically thin 
and show a more similar decrease.

The good agreement between the ALMA data and the $df = 320$ model supports the validity of 
the approach taken here, providing a better fit to emission from the optically thinner CO isotopologues. 
This agreement suggests that the CO column density exhibits a sharp drop across the CO snowline, 
consistent with the presence of a thick VIRaM layer near the midplane of this disk. 
The sharp transition in this model provides a more accurate representation of the physical conditions.

Both the \citet{Qi+2011, Qi+2015} and \citet{Zhang+2021} modeling approaches rely on fitting the 
spectral energy distribution (SED), resolved millimeter dust continuum data, and multiple CO isotopologue emission lines 
in a consistent framework. However, the fundamental difference between these studies lies in how the 
vertical temperature structure near the disk midplane is modeled.

We employed the {\em DIAD} model adopted from \citet{Qi+2011, Qi+2015}, which directly addresses the degeneracy in vertical temperature structure that arises when models are constrained by  dust emission (e.g., the SED). The {\em DIAD} model accounts for the concentration of large grains near the midplane, enhancing the temperature gradient between the surface and interior layers. This configuration can result in a vertically isothermal region (VIRaM layer) near the midplane of some disks, influencing the CO distribution and its emission properties.
In contrast, the \citet{Zhang+2021} analysis uses thermal-chemical models, such as RAC2D, to derive the dust temperature structure. While these models provide detailed chemical insights, they do not explicitly capture the vertically isothermal structure near the midplane. The difference in modeling approaches contributes to the different interpretations 
of the CO snowline transition in the HD~163296 disk across studies.

\subsection{Effect on Disk Mass Estimates}
The emission from rare CO isotopologues remains one of the best methods for estimating total disk gas masses, a critical parameter for understanding many aspects of planet formation \citep[e.g.,][]{Bergin+2017,Miotello+2023}. However, accurate disk mass estimates are challenging due to factors such as isotope-selective CO photodissociation, CO freeze-out, and the complex thermal structure of disks. Low CO masses observed in T Tauri disks \citep{Ansdell+2018} are often attributed to chemical reprocessing of CO into more complex ices \citep{Bosman+2018} or the sequestration of CO onto larger bodies \citep{Krijt+2018}. These mechanisms are less significant in warmer Herbig Ae disks \citep{Kama+2020}, where CO freeze-out is inefficient. However, discrepancies in disk mass estimates may also arise from structural differences within the disks.

As the models in Section~\ref{sec:toy} demonstrate, the thickness of the VIRaM layer plays a critical role in shaping the vertical temperature distribution, which directly impacts the CO column density's radial variation across the CO snowline. This, in turn, influences the emission of CO isotopologues and affects the derived disk mass.

\begin{figure*}
\centering
\includegraphics[scale=0.8]{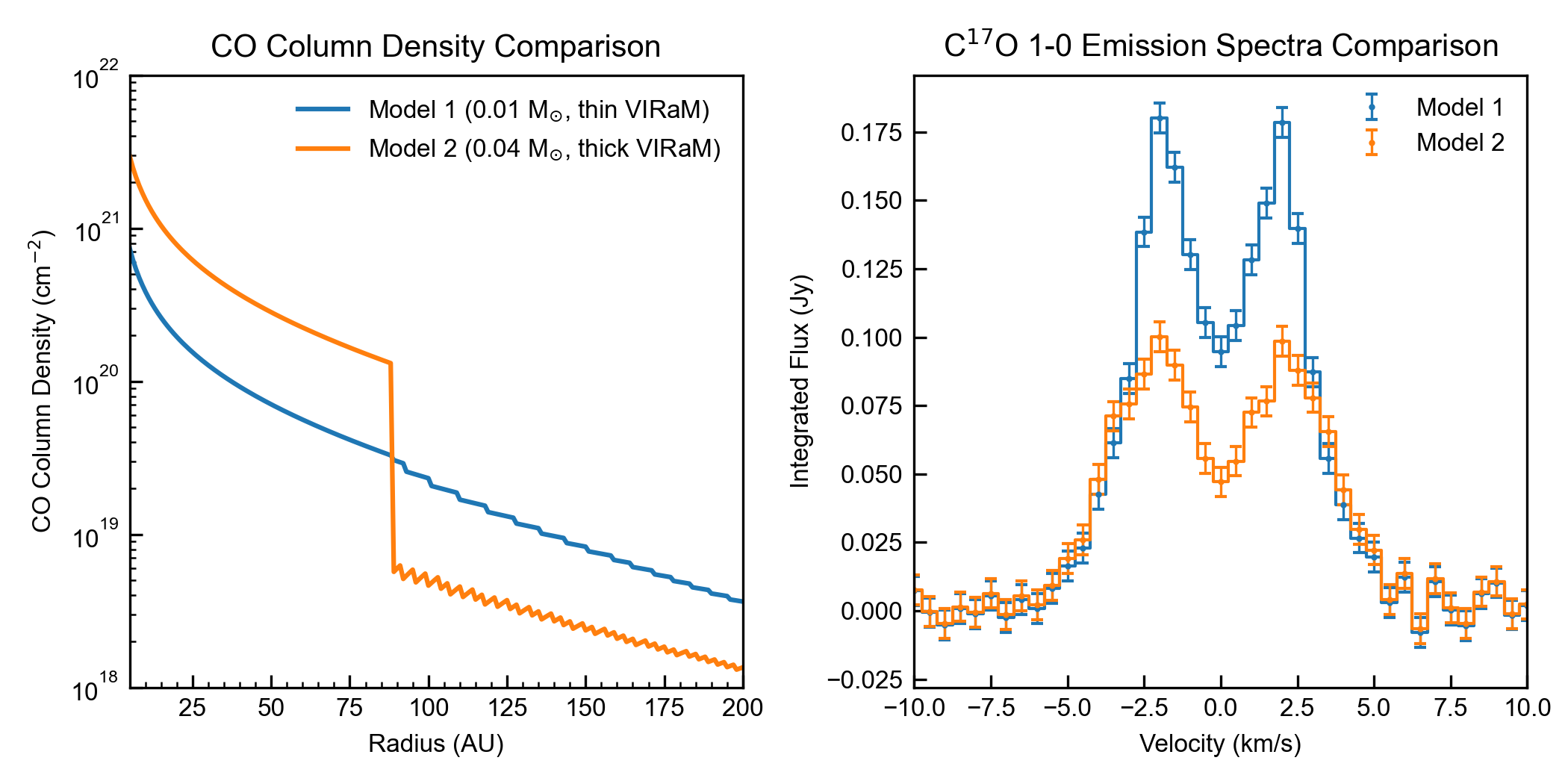}
\caption{
Comparison of two disk models with different VIRaM layer thicknesses and disk masses. 
The left panel shows the CO column density profiles for Model 1 (thin VIRaM layer, 0.01 M$_\odot$, blue) and 
Model 2 (thick VIRaM layer, 0.04 M$_\odot$, orange). Model 2 exhibits a sharp drop in the CO column density across the CO snowline, 
while Model 1 shows a more gradual decline. The right panel shows C\(^{17}\)O 1-0 emission spectra for both models. 
The integrated flux of Model 1 is consistently higher than Model 2, despite Model 2 having higher CO column density inside the snowline,
as emission from beyond the CO snowline, where the gas is optically thin, dominates at this (low) angular resolution.
\label{fig:diskmass}}
\end{figure*}

To illustrate the impact of disk thermal structure on mass inferences, we compared two models: 
Model 1, with a thin VIRaM layer and a disk mass of 0.01 M$_\odot$, and 
Model 2, with a thick VIRaM layer and a disk mass of 0.04 M$_\odot$. 
Figure~\ref{fig:diskmass} (left panel) shows the CO column density profiles for both models. 
In Model 2, the CO column density inside the CO snowline is higher than in Model 1, due to the larger disk mass (by a factor of 4). However, beyond the CO snowline, the CO column density in Model 2 drops sharply, resulting in a lower CO column density compared to Model 1. Figure~\ref{fig:diskmass} (right panel) presents the C\(^{17}\)O 1-0 spectra for these models at $1\farcs5$ resolution, generated using the \texttt{integrated\_spectrum} function from the Python package \texttt{GoFish} \citep{Teague2019}. Despite Model 2 having a higher CO column density inside the CO snowline, its spectrum is lower than that of Model 1 because the emission beyond the CO snowline dominates the integrated flux in both models. In Model 2, the emission inside the CO snowline becomes optically thick, reducing its sensitivity to CO column density variations.

This analysis highlights how the thickness of the VIRaM layer significantly sharpens the CO snowline transition and alters CO isotopologue emission. If the detailed disk structure, particularly the VIRaM layer thickness, is not properly accounted for, disk mass estimates derived from CO isotopologue emission may be substantially inaccurate.

\citet{Williams+2014} used a grid of simple parametric disk models to argue that total disk gas masses could be reasonably estimated from unresolved integrated emissions of \(\mathrm{^{13}CO}\) and \(\mathrm{C^{18}O}\). Subsequent studies, such as \citet{Miotello+2016}, incorporated more detailed chemistry and demonstrated that uncertainties of up to an order of magnitude could arise, especially for disks with masses greater than 0.001 M$_\odot$, due to optical depth effects. Our results extend these findings by showing that even the emission from low-abundance tracers like \(\mathrm{C^{17}O}\) can be significantly influenced by assumptions about the disk’s physical conditions. Accounting for complex thermal structures, including the thickness of the VIRaM layer, is crucial for accurately interpreting CO isotopologue emissions and deriving disk masses.

To reliably detect a dip in the derivative over the CO snowline for the thick VIRaM case, three observational criteria must be met:  
(1) The tracer's line optical depth must remain optically thin close within the CO snowline, ensuring sensitivity to the local column density changes;  
(2) The instrument must have sufficient sensitivity to detect the tracer emission even after the column density drops by more than an order of magnitude beyond the CO snowline;  
(3) The spatial resolution of the instrument must be high enough to resolve the sharp transition at the CO snowline, preventing the profile from being smoothed out by the beam size.

\subsection{Prospects with ngVLA}
An important implication of our modeling is that high-resolution observations of 
optically thin CO isotopologue line emission are needed to locate CO snowlines in disks 
with different thermal structures and to confront associated predictions. 
The combination of resolution and sensitivity will become even more important for disks 
around pre-main-sequence stars of lower luminosity than HD~163296 -- the T Tauri stars --  where the 
CO snowline is located closer to the star, typically $\sim20 - 30$~au, and CO freeze-out is more significant. 
While it is clear that ALMA can effectively probe optically thin CO isotopologue lines in 
a large, warm, massive disk like the one around the Herbig Ae star HD~163296, even the relatively deep 
($2-3$ hr) ALMA MAPS observations of disks around T Tauri stars resulted in tentative or 
modest detections of C$^{17}$O 1-0 emission at $\sim0\farcs25$ resolution \citep{Oberg+2021}. 
Substantially better observational capabilities than currently available with ALMA 
will be needed to obtain sufficient signal-to-noise ratio
to extend the study of CO snowlines to a broader population of disks. 

For sharp CO snowlines, detecting radial C/O variations becomes easier, offering a clearer understanding of 
the abundance and location of molecules in disks and their role in the formation of planetesimals and icy planets. 
The ngVLA is envisioned to improve ALMA's sensitivity by up to an order of magnitude 
for observations in the frequency range $35-116$~GHz where ngVLA and ALMA overlap.  
To date, most interest in ngVLA for protoplanetary disks 
has focused on milliarcsecond-resolution observations of thermal dust emission to image terrestrial planet formation
zones of disks, to reveal substructures associated with young planets \citep[e.g.,][]{Ricci+2018}. 
However, the upper end of the ngVLA frequencies includes the ground-state transitions of the suite of 
rare CO isotopologues that can provide optically thin tracers for a wide range of CO column densities
at angular resolutions appropriate for CO snowlines in the large population of T Tauri stars located at the 
$\sim140$~pc distance of nearby star-forming regions. Deep, resolved ngVLA observations to exploit 
these diagnostics. For example, the ALMA MAPS detection of C$^{17}$O 1-0 in GM Aur at 
the $3.5\sigma$ level suggests the ngVLA could provide a resolved radial profile at sufficient 
signal-to-noise ratio to reveal the CO snowline location with a comparable investment of observing time. 
Observations of a large sample have the potential to provide a much more nuanced understanding of 
the factors that contribute to CO isotopologue emission from disks and the effects on inferred physical 
properties such as disk gas masses. 

\section{Conclusions}
\label{sec:conclusion}
We use simple models of protoplanetary disks to demonstrate the importance of accounting for the 
disk thermal structure on the CO column density distribution, in particular the vertical extent of an isothermal 
region at the midplane, and we investigate the implications for observations of low-$J$ rotational lines 
of CO isotopologues at millimeter wavelengths. The main findings are:  

\begin{itemize}

\item A thick vertical isothermal layer produces a sharp drop in CO column density across the CO snowline.
Simulated CO isotopologue images from the ngVLA (for J=1--0 lines) and ALMA (for J=2--1) lines 
demonstrate the importance of low line optical depth in revealing the location of 
the sharp CO snowline transition. We show that a local minimum in the derivative of the radial 
emission profile can serve to identify the CO snowline location.

\item Applying the concepts from the models to archival ALMA observations of CO isotopologue emission 
from the Herbig Ae star HD~163296 system reveals a sharp CO snowline transition, with the CO column density decreasing by more than a factor of 20 between 70 and 90~au. This provides clear evidence of a distinct CO snowline near 80~au, consistent with the location previously inferred from chemical signatures of CO depletion. Our analysis also refines the constraints on the CO depletion fraction.

\item The presence of a sharp CO snowline has implications on total disk gas estimates derived
from CO isotopologue emission. Substantial, systematic, underestimates of the disk gas mass can result
from this different thermal structure, even when the depletion due to CO freeze-out is considered. 

\end{itemize}

The development of the ngVLA, which will provide a substantial sensitivity improvement 
over ALMA for high-resolution observations of the J=1--0 lines of CO isotopologues, will enable 
extending this approach to CO snowline studies to many more systems, in particular for 
a sample of the large population of T Tauri disks in nearby star-forming regions. 

\begin{acknowledgments}
We thank the anonymous referee for a constructive report, which improved the paper. C.Q. thanks K. {\"O}berg, S. Andrews, and C. Espaillat for helpful discussions about the results presented in this paper.
C.Q. acknowledges support for this work from the NRAO ngVLA Community Study program. 
This paper makes use of the following ALMA data: ADS/JAO.ALMA\#2016.1.00884.S and ADS/JAO.ALMA\#2018.1.01055.L. ALMA is a partnership of ESO (representing its member states), NSF (USA), and NINS (Japan), together with NRC (Canada), NSTC and ASIAA (Taiwan), and KASI (Republic of Korea), in cooperation with the Republic of Chile. The Joint ALMA Observatory is operated by ESO, AUI/NRAO and NAOJ.
The National Radio Astronomy Observatory is a facility of the National 
Science Foundation operated under a cooperative agreement by Associated Universities, Inc.
\end{acknowledgments}

\vspace{5mm}
\facilities{ALMA}

\software{Astropy \citep{Astropy2013,Astropy2018,Astropy2022}, \code{bettermoments} \citep{Teague+2018}, \code{CASA} \citep{CASA+2022},  \code{GoFish} \citep{Teague2019}, Matplotlib \citep{Hunter2007}, NumPy \citep{vanderWalt2011}, RADEX \citep{vanderTak+2007}, SciPy \citep{Virtanen2020}, vis\textunderscore sample \citep{Loomis+2018}.}

\clearpage

\bibliography{ngvlastudy.bib}{}
\bibliographystyle{aasjournal}

\end{document}